\documentclass[12pt]{article}
\usepackage{amsmath,amssymb,graphicx,color}

\newcommand{\bea}{\begin{eqnarray}}
\newcommand{\eea}{\end{eqnarray}}
\newcommand{\be}{\begin{equation}}
\newcommand{\ee}{\end{equation}}
\newcommand{\vs}[1]{\vspace{#1 mm}}

\newcommand{\dsl}{\pa \kern-0.5em /}

\newcommand{\pa}{\partial}

\newcommand{\nn}{\nonumber\\}

\begin{document}
\topmargin 0mm
\oddsidemargin 0mm

\begin{flushright}

USTC-ICTS/PCFT-25-08\\

\end{flushright}

\vspace{2mm}

\begin{center}

{\Large \bf Branes in String/M-Theory\footnote{Contribution to the proceedings of the 3rd summer school on strings, fields and holography in Nanjing (2023).}}

\vs{10}

{\large J. X. Lu}

\vspace{4mm}

{\em
Interdisciplinary Center for Theoretical Study\\
 University of Science and Technology of China, Hefei, Anhui
 230026, China\\
 \medskip
 Peng Huanwu Center for Fundamental Theory, Hefei, Anhui 230026, China\\
 %\vs{4}
}

\end{center}

\vs{10}

\begin{abstract}
This is the writeup of lectures delivered in Asian Pacific introductory school on superstring and the related topics in Beijing (2006) and the expanded version of these lectures in the 3rd summer school on strings, fields and holography in Nanjing (2023). It intends to give a historical as well as a pedagogical account of the development in finding the 1/2 BPS extended string solitons in the early stage of the so-called second string revolution before which those objects were thought to be unrelated to strings. Non-susy solutions which are related to brane/anti brane systems or non-BPS systems are also discussed.
\end{abstract}

\section{ Overview/Motivation}

There occurred two revolutions in the course of superstring development. In the so-called first superstring revolution (1984-1985),  it was established that there exist five perturbatively consistent quantum superstring theories, namely, Type IIA, Type IIB, Type I, Heterotic SO(32) and Heterotic $E_{8} \times E_{8}$, each of which requires 10 spacetime dimensions (nine spaces and one time) and spacetime supersymmetry. In the so-called second string revolution, there discovered many non-perturbative states nowadays called p-branes or NS(-NS) p branes and Dp branes, which played an important role in such revolution, giving rise to various dualities and revealing the existence of a unique but not-yet completely established unified theory called M-Theory (See \cite{Witten:1997fz} for example).

The M-theory has its maximal eleven-dimensional spacetime and unifies not only the five known 10 dimensional perturbative string theories but also the once isolated 11 dimensional supergravity as six different limits, as shown in 
Figure 1.

\begin{figure}[!hbp]
\begin{center}
\includegraphics[scale= 0.4]{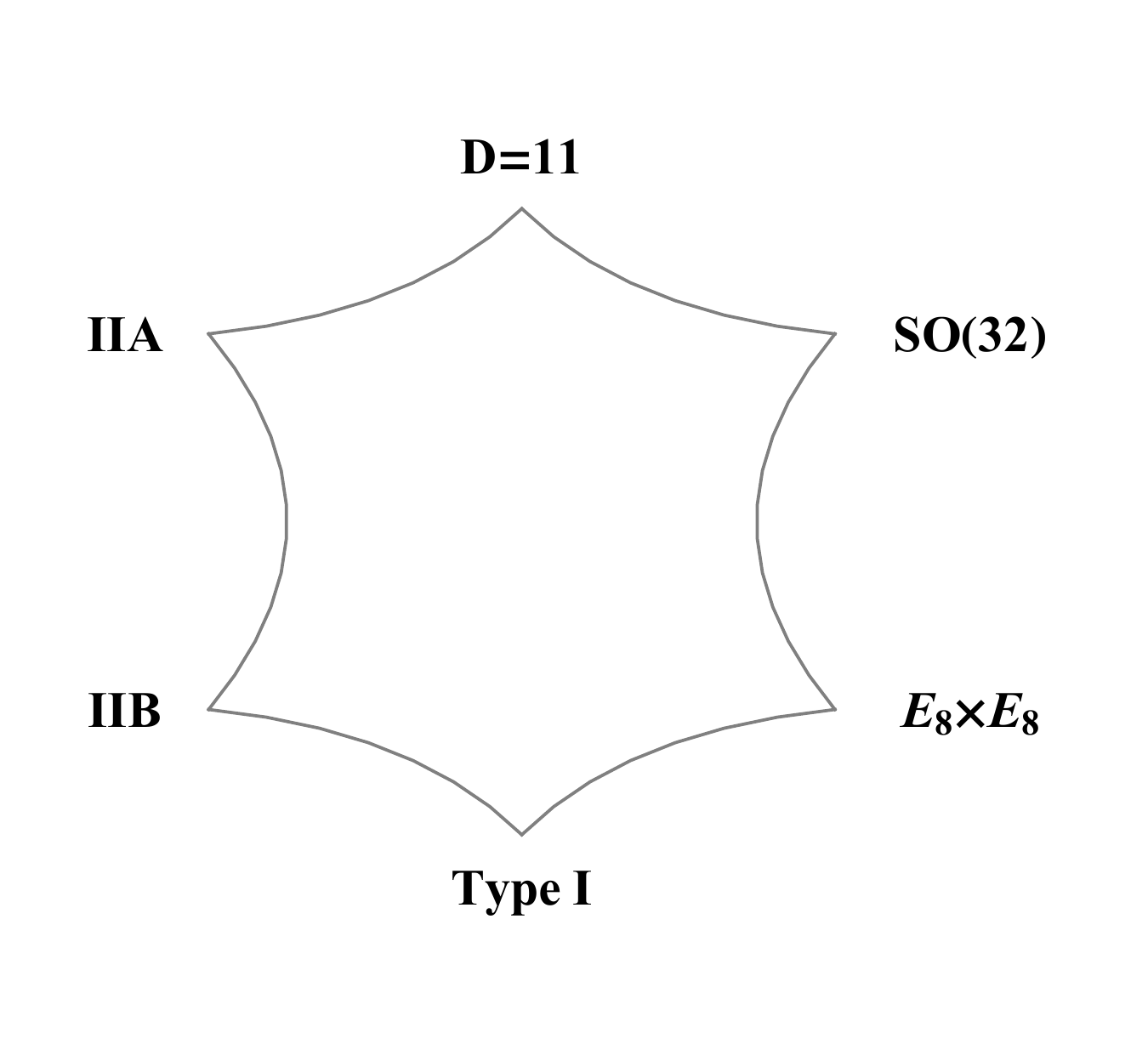}
\end{center}
\caption{M-Theory}
\end{figure}

 It also answers many puzzles left after the first string revolution such as the so-called ``embarrassing rich problem'' (too many string theories and one real world) and the status of eleven dimensional supergravity. 
In particular, the Dp branes have a dual description of either closed string or open string which provides the basis for the AdS/CFT as well as the Matrix Theory proposal for M-Theory.

\medskip

\noindent
{\bf The First Superstring Revolution:}

\medskip

 In the early days of superstrings and supermembranes,  two views were taken in the world of quantum gravity and grand unified theory. 
 
 People in string community (mainly in US) then were very against the study of supermembranes, i.e., the extended objects with their spatial dimensionality higher than one, for the simple reason that only strings as a (1 + 1)-dimensional conformal field theory (CFT) can potentially be first quantized due to enough underlying local symmetries, in particular, the Weyl symmetry,  which can be used to set the worldsheet flat in a given coordinate patch, implying no worldsheet physical propagating gravity or the worldsheet propagating gravity decoupled. 

 This was reflected, for example, in the first well-known ``Superstring Theory'' textbook by Green, Schwarz and Witten with the quotation \cite{GSW}``Weyl invariance, or at least the ability to locally gauge away the $h_{\alpha\beta}$ dependence, is central in the physics of strings. This is one of the things that singles out strings as opposed to, say, membranes. Membranes and objects of still higher dimensionality have another glaring problem, as follows.  Equation (\ref{p-action}) (i.e., the p-brane Polyakov-type action) defines an (p + 1)-dimensional quantum field theory, which is by power counting  renormalizable for $p = 1$ and non-renormalizable for $p > 1$. Making sense of (\ref{p-action}) as a quantum theory for $p > 1$ is as difficult a problem as making sense of general relativity as a quantum theory. Thus, membranes  or higher dimensional objects would be hardly be a promising start toward quantum gravity''.  
 \be
 \label{p-action}
 S_{p} = - \frac{T_{p}}{2} \int d^{p + 1} \sigma \sqrt{- h} \left(h^{\alpha\beta} \partial_{\alpha} X^{\mu} \partial_{\beta} X^{\nu} g_{\mu\nu} (X) - (p - 1)\right) .
 \ee

While those in Europe, mostly in England,  took a different view by asking if people are interested in strings, why not higher-dimensional extended objects for quantum gravity. There are some rationals behind this view.

When Green and Schwarz (GS) made use of the so-called fermionic local $\kappa$-symmetry, discovered by Warren Siegel from the supersymmetric particle action \cite{Siegel:1983hh}, to give rise to Type I and Type II superstring theories with manifest spacetime supersymmetries without the need of Gliozzi-Scherk-Olive (GSO) projection, called the Green-Schwarz formalism of superstring theories, certain $\Gamma$-matrix identity must hold which can be true in spacetime dimension $D = 10, 6, 4$ and $3$, corresponding to those dimensions for which the super Yang-Mills theories exist.  This led to the belief that this $\kappa$-symmetry would be hardly possible for the worldvolume actions of objects with their spatial dimensionality higher than one, possessing their respective manifest spacetime supersymmetries.

It was the late Polchinski and his collaborators who evaded this by showing explicitly that this local fermionic symmetry can be used to construct the supermembrane (actually a super 3-brane) action in six spacetime dimensions \cite{Hughes:1986fa}.

Shortly after this and following the same procedure, Bergshoeff, Sezgin and Townsend \cite{Bergshoeff:1987cm} found corresponding actions for other values of $d$ and $D$, called super p-branes where $p = d - 1$ is the number of spatial dimensions of the worldvolume (here $D$ stands for the spacetime dimensions).  For example, for the 11 dimensional supermembrane action,  they showed that the $\kappa$-symmetry itself requires that 11 dimensional supergravity must be on-shell, i.e., the equations of motion (EOMs) of 11 dimensional supergravity hold, when the supermembrane couples with the supergravity. 

This, to some extent, hints that the 11 dimensional supergravity multiplet appears to give rise to the massless modes of supermembrane if the latter could be quantized.

Soon after this,  the Polyakov-type actions for a large class of super p-branes in diverse dimensions had been classified \cite{Achucarro:1987nc}. Each of their actions needs the $\kappa$-symmetry and this symmetry can hold only if the corresponding supergravity fields satisfy certain constraints consistent with their EOMs when the p-brane is coupled with the supergravity background.

Moreover, Duff, Howe, Inami and Stelle \cite{Duff:1987bx} showed how the action for a $(p - 1)$-brane in $(D - 1)$-dimensions could be derived from that for a $p$-brane in $D$-dimensions via the so-called double-dimensional reduction. In particular,  the Type IIA superstring action in 10 dimensions can be obtained from the supermembrane action in 11 dimensions.

Precisely because of these developments, it gave a surge of interest in super p-branes, in particular, the 11 dimensional supermembrane, even though these higher dimensional objects do not appear to be quantizable.

\begin{figure}[!hbp]
\begin{center}
\includegraphics[scale= 1.0]{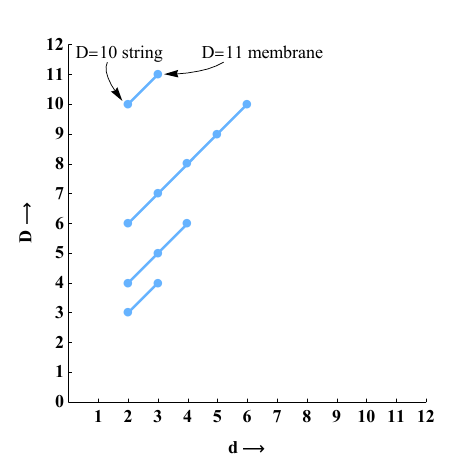}
\end{center}
\caption{The old brane scan}\label{oldbs}
\end{figure}

 Each of these super p-branes considers  only its worldvolume scalar supermultiplet, i.e., consisting of  only the worldvolume scalars and spinors in the multiplet, and as mentioned above they had been classified in diverse dimensions in \cite{Achucarro:1987nc}. According to this classification, Type II p-branes, i.e., those with $N = 2$ spacetime supersymmetry, do not exist for $p > 1$, which can be summarized in the so-called old brane scan \cite{Duff:1992hu} in Figure~\ref{oldbs}.
 
Given the above, there exist two puzzles remained. If the 10 dimensional superstrings are the whole story,  how to explain and to understand the 11 dimensional supergravity, noting that the dimensional reduction of 11 dimensional supergravity gives rise to  Type IIA supergravity.  In addition, as mentioned above, Duff, Howe, Inami and Stell \cite{Duff:1987bx} demonstrated that the Type IIA superstring action can be obtained from the eleven dimensional supermembrane action via the so-called double-dimensional reduction.

 In other words, the 11 dimensional supermembrane appears to be more fundamental than the superstring.  If the Type IIA superstring can be quantized plus its connection to the 11 dimensional supermembrane, it hints that there should exist a quantum theory of the 11 dimensional supermembrane.  This opens up people's interests in
 seeking how to quantize the supermembrane.

In these efforts, the Matrix regularization procedure stands out. Its basic ideas are: Using the three diffeomorphsims of M2 brane worldvolume,  one can set its worldvolume metric $h_{0a} = 0, \, h_{00} \propto - \det h_{ab}$ with $a, b = 1, 2$.  In addition,  the light-cone gauge of $X^{+} \propto \tau$ is taken.

With the above gauge choices, the residue symmetries of M2 worldvoume are the diffeomeorphsims preserving the area of the M2. As such, any function defined on the M2 worldvolume can be represented by a U(N) matrix with $N \to \infty$, and the dynamics of M2 can then be described by the following Hamiltonian

\be
H = \frac{1}{2\pi l^{3}_{11}} {\rm Tr}\left(\frac{1}{2} \dot X^{i} \dot X^{i} - \frac{1}{4}\left[X^{i}, X^{j}\right] \left[X^{i}, X^{j}\right] + \frac{1}{2} \theta^{T} \gamma^{i} \left[X^{i}, \theta\right]\right),
\ee
where $l_{11}$ is the 11 dimensional Planck length, $X^{i}$ ($i = 1, 2, \cdots 9$) and the 9-dimensional Majorana spinor $\theta$ are all $N\times N$ matrices.

By this, we convert the (1 + 2)-dimensional M2 brane dynamics to a (1 + 0)-dimensional infinite Matrix quantum mechanics which appears to be quantizable \cite{deWit:1988wri}, therefore providing the hope to quantize the supermembrane. 

However, not long after this, de Wit, Luscher and Nicolai \cite{deWit:1988xki}  showed that the energy spectrum of this system is continuous, suggesting the instability of this system, therefore ending  people's further interest of studying the 11D supermembrane.

One has to wait about 10 years, when the non-perturbative effects of superstring are considered and become important, to realize the physical significance of this continuous spectrum. 

The would-be first quantized Matrix theory turns out to be a second quantized one, containing multi-particle and multi brane states, therefore the continuous spectrum problem is solved and the resulting Matrix theory provides a concrete proposal for M-theory\cite{Banks:1996vh} (See also the nice review article \cite{Taylor:2001vb} by Taylor).

In addition to the above, by the end of the first superstring revolution, we have encountered other problems such as
the so-called ``embarrassing rich problem''(Too many theories but one real world), experimental testing problem\footnote{Though the Calabi-Yau compactification of the heterotic $E_{8} \times E_{8}$ superstring \cite{Candelas:1985en}  does give some thing which looks like the particle physics standard model with one supersymmetry, there exist also other light modes such as various scalar multiplets which are not seen in the real world. In the perturbative region of superstrings, their contribution in low energy cannot be eliminated.  In addition, picking up this special string theory is due to a phenomenological preference, not by the theory itself or a more fundamental reason.}. All the five superstring theories are only perturbatively well-defined. In other words, they are only asymptotically well-defined.  A well-defined unified theory cannot be an asymptotically one since its coupling, vacuum structures, except for its fundamental inputs such as its tensions and other fundamental constants as well as some initial or boundary conditions, should all be determined dynamically. 

Addressing any of these issues needs to go to the non-perturbative region of superstrings, i.e., $g_{s} \sim {\cal O} (1)$.

The worldsheet action of superstring can be used to study its perturbative behaviors. How to study its non-perturbative effects was a challenging problem. We didn't have a non-perturbative formulation of a given superstring theory and the only thing available then was the corresponding 10 dimensional supergravity which was believed to be the low energy effective theory of the perturbative superstring\footnote{The superstring field theory may be good in providing such a non-perturbative description but its development, even nowadays, is still in its infancy.}.

This is due to that for each of the five perturbative superstring theories, its massless spectrum corresponds to the corresponding supergravity supermultiplet plus possible super Yang-Mills multiplet (such as in type I and two heterotic string theories).

 For example for IIA theory, the massless spectrum can be given as the tensor product of left and right movers with their 8-spinors having opposite chirality as
 \be
(8_{v} \oplus 8_{s}) \otimes (8_{v} \oplus 8_{c}) = 8_{v}\otimes 8_{v} \oplus 8_{v}\otimes 8_{c} \oplus 8_{s}\otimes 8_{v} \oplus 8_{s}\otimes 8_{c}.
\ee
Here the bosonic Neveu-Schwarz $-$ Neveu-Schwarz (NSNS) sector gives the gravity multiplet and decomposes according to the little group SO(8) as
\be
8_{v} \otimes 8_{v} = 1 \oplus 28 \oplus 35 = \phi \oplus B_{ij} \oplus g_{ij},
\ee
where $\phi$ is a singlet (the dilaton),  $B_{ij}$ a 2-form antisymmetric tensor (the Kalb-Ramond field) and $g_{ij}$ the traceless symmetric tensor (the graviton) all under SO(8).  While the so-called Ramond$-$Ramond (RR) sector gives the additional bosonic form potentials from the bi-linear fermionic fields with opposite chirality as
\be
8_{s} \otimes 8_{c} = 8 \oplus 56 = A_{i} \oplus A_{ijk},
\ee
where $A_{i}$ is a one-form vector while $A_{ijk}$ is a 3-form tensor.  Concretely, we have
\be
 \psi_{a} \tilde\psi_{\dot a} = \gamma^{i}_{a \dot a} A_{i}  + \frac{1}{3!} \gamma^{ijk}_{a \dot a} A_{ijk}.
 \ee
 For Type IIB superstring, its massless spectrum comes from the tensor product of left and right movers with the fermions having the same chirality as
\be
 (8_{v} \oplus 8_{s}) \otimes (8_{v} \oplus 8_{s}) = 8_{v}\otimes 8_{v} \oplus 8_{v} \otimes 8_{s} \oplus 8_{s} \otimes 8_{v} \oplus 8_{s}\otimes 8_{s}.
 \ee
 The bosonic NSNS sector remains the same as in Type IIA case while the bosonic RR sector gives also additional bosonic form potentials as
 \be
 8_{s} \otimes 8_{s} = 1 \oplus 28 \oplus 35_{+} = \chi \oplus A_{ij} \oplus A^{+}_{ijkl},
 \ee
 where $\chi$ a zero-form scalar, an axion-like field, $A_{ij}$ is a 2-form tensor and $A^{+}_{ijkl}$ is a 4-form tensor, satisfying the following self-duality
 \be
 A^{+}_{i_{1}i_{2}i_{3}i_{4}} = \frac{1}{4!} \epsilon_{i_{1} i_{2} i_{3} i_{4}}\,^{i_{5}i_{6}i_{7} i_{8}} A^{+}_{i_{5} i_{6} i_{7} i_{8}},
 \ee
 which reduces its degrees of freedom (DOF) by half. Concretely, we have
 \be
 \psi_{a} \tilde\psi_{b} = (\gamma^{9})_{ab} \chi + \frac{1}{2!} (\gamma^{ij})_{ab} A_{ij}  +  \frac{1}{4!} (\gamma^{ijkl})_{ab} A^{(+)}_{ijkl}.
 \ee
In the above, $\gamma^{i}$ with $i = 1, 2, \cdots 8$ are the SO(8) Dirac matrices and $\gamma^{9} \equiv \gamma^{1} \gamma^{2} \cdots \gamma^{8}$ the eight dimensional chiral operator. 

The NSNS 2-form potential $B_{2}$ appears in any (except for Type I) consistent superstring theory and is always with the gravity multiplet, i.e., the NSNS sector.  Its appearance is completely expected since the string carries the so-called NSNS charge and as a one-dimensional extended object, it must couple with a 2-form potential just like a point-charge must couple with a U(1) 1-form potential, given what we have discussed early.

The RR form potentials in either IIA or IIB come from the bi-linear spinors and are hard to understand their origins from the perturbative string perspective though it is clear that the NSNS 2-form potential $B_{2}$ couples to the underlying fundamental string. Even so, it is however that such a natural question of what is the magnetic dual of a string in 10 dimensions, which is supposed to be a NSNS 5-brane, had never been asked until the very end of 1980's when a very few including myself started seriously to  address the non-perturbative issues regarding how strings are related to other higher dimensional objects. 

Michael Duff \cite{Duff:1987qa} was the first to notice that there may exist a duality between the heterotic string with either $SO(32)$ or $E_{8} \times E_{8}$ and the corresponding so-called heterotic five-brane, therefore conjectured the existence of heterotic five-brane,  based on the observation among other things that there exist two equivalent dual formulations of N = 1 supergravity plus the respective super Yang-Mills in 10 dimensions, one with the NSNS 3-form field strength or 2-form NSNS potential \cite{Chapline:1982ww}, therefore associated with the heterotic string, while the other with the NSNS 7-form field strength or 6-form potential \cite{Chamseddine:1981ez} associated with the so-called heterotic five-brane. The first non-trivial evidence in support of this was made by Strominger \cite{Strominger:1990et} in finding the heterotic five-brane with its core as an instanton from the low energy theory of heterotic string as a solution which preserves one half of spacetime supersymmetry\footnote{The heterotic string solution was also later found in \cite{Duff:1990wu} from the dual formulation of \cite{Chamseddine:1981ez} when the relevant higher order corrections are considered.}.  Subsequently, Duff and myself \cite{Duff:1990wv} found the so-called elementary five-brane solution from the 10 dimensional  N = 1 supergravity which preserves also 1/2 spacetime supersymmetry and was shown to correspond to the zero size instanton  limit of Strominger's solution. Moreover, this five-brane solution solves the respective equations of motion of  all the 10 dimensional supergravities and preserves the respective 1/2 spacetime supersymmetries, therefore as a 1/2 BPS five-brane solution of the respective supergravities.  All these provide further evidence in support of the existence of NSNS five-branes.

The discovery of various supergravities was actually a few years earlier than the corresponding perturbative superstrings. They are based on the representations of the underlying supersymmetry algebra and the spacetime localization of the corresponding supersymmetry (SUSY) transformations. Since the algebra itself has nothing to do with the string coupling, each of the supergravities, whose concrete form, is related to the relevant low energy scale (the corresponding Planck scale) as in every effective theory description\footnote{If the on-shell bosonic and fermionic degrees of freedom remain the same but the underling concrete description depends on the respective low energy scale (there can be more than one low energy scale, the concrete form of the supergravity can also be more than one), we view such supergravity or supergravities as the same supergravity theory and this is consistent with our current understanding. For example, we view the usual 11 dimensional supergravity and 10 dimensional Type IIA supergravity as two different forms of the same supergravity theory, each of which corresponds to their respective effective description in the respective low energy scale. A rather detail discussion of this is given in \cite{lu1}.}  and as such each of these should be viewed as the low energy effective theory of the underlying non-perturbative string/M theory, rather than, as previously thought, that  of the perturbative one, see a discussion of this in \cite{lu1}.

\medskip

\noindent
{\bf The Second Superstring Revolution}
\medskip

In other words, if supergavities are the respective low energy effective theories of the underlying non-perturbative superstrings (independent of the underlying string coupling or valid for any string coupling), we can simply ignore the perturbative picture regarding the RR potentials (named in the perturbative sense) discussed earlier and naturally associate each of them with the corresponding branes, just like a point charge coupled with a 1-form potential and a one-dimensional charged string coupled with a 2-form potential,  and in general a (1 + p)-form potential coupled with a p-brane which carries the corresponding charge\footnote{This is also consistent with our current understanding regarding the absence of global symmetries including higher form symmetries in any consistent quantum gravity theory. In other words, string theories as consistent quantum gravity theories imply the existence of these dynamical extended objects.}. In other words, we have in general the following:
 \bea
 A_{1} &=& A_{\mu} d x^{\mu} \qquad\qquad \to \qquad\qquad \qquad \quad {\rm coupled\, with\, a\, charged\, point\, particle},\nn
 A_{2} &=& \frac{1}{2!} A_{\mu\nu} d x^{\mu} \wedge d x^{\nu} \qquad \to \qquad \qquad \quad {\rm coupled \,with\, a\, charged\, one\,brane},\nn
 A_{3} &=& \frac{1}{3!} A_{\mu\nu\rho} d x^{\mu}\wedge d x^{\nu} \wedge d x^{\rho} \qquad\to \quad \quad {\rm coupled \,with\, a\, charged\, two\, brane},\nn
 &\,&\vdots\nn
 A_{1 + p} &=& \frac{1}{(1 + p)!} A_{\mu \cdots \sigma} d x^{\mu} \wedge \cdots \wedge d x^{\sigma} \quad \to \quad  {\rm coupled \,with\, a\, charged\, p\, brane}.\nn
 \eea

To demonstrate the correctness of the above, we need to show indeed the existence of these branes associated with the form potentials in various supergravities along with the fundamental string which is associated with the NSNS 2-form $B_{2}$ by finding their solutions from these supergravities and at the same time to show that their existence is independent of the string coupling.

Duff and myself were among a few to start such a journey to find the brane solutions or string solitons preserving one half of spacetime SUSY, called 1/2 BPS branes. For examples, Duff and I found the so-called the elementary 5-brane (i.e., the NSNS 5-brane) \cite{Duff:1990wv} mentioned earlier, the seld-dual superthreebrane (the D3 brane) \cite{Duff:1991pea} and the general 1/2 BPS p-branes in diverse dimensions \cite{Duff:1993ye}.  Note that the fundamental string or F-string is also such a 1/2 BPS state found earlier \cite{Dabholkar:1990yf}. See also discussions in \cite{Callan:1991ky} for 1/2 BPS NSNS 5-brane solitons and in particular their zero modes in Type II superstrings. The black p-brane solutions in 10 dimensions were also found earlier in \cite{Horowitz:1991cd}.

Though these 1/2 BPS p-branes are found as solutions of various supergravities, their existence is independent of the underlying string coupling and they are fundamental dynamical objects of non-perturbative string/M theory.  This is due to that their respective  ADM mass per unit brane volume or their  tension \cite{Lu:1993vt} equaling to their corresponding charge, i.e., $M_{p} = Q_{p}$ in certain units, i.e., the BPS property, is protected by the underlying unbroken SUSY and the quantized charge.  In other words, this relation is exact and is independent of the underlying string coupling or is good for any string coupling\footnote{Though the solutions themselves may be corrected when higher order corrections to the low energy effective theory are included.}.   For example, the existence of these 1/2 BPS objects can also be deduced purely from their respective supersymmetry algebra with the proper central extension \cite{Townsend:1995gp,Townsend:1996xj, Townsend:1997wg}  and this also shows that these objects are the fundamental objects in the underlying non-perturbative theory following \cite{Witten:1978mh}. 

\begin{figure}[t]
\begin{center}
\includegraphics[scale= 0.4]{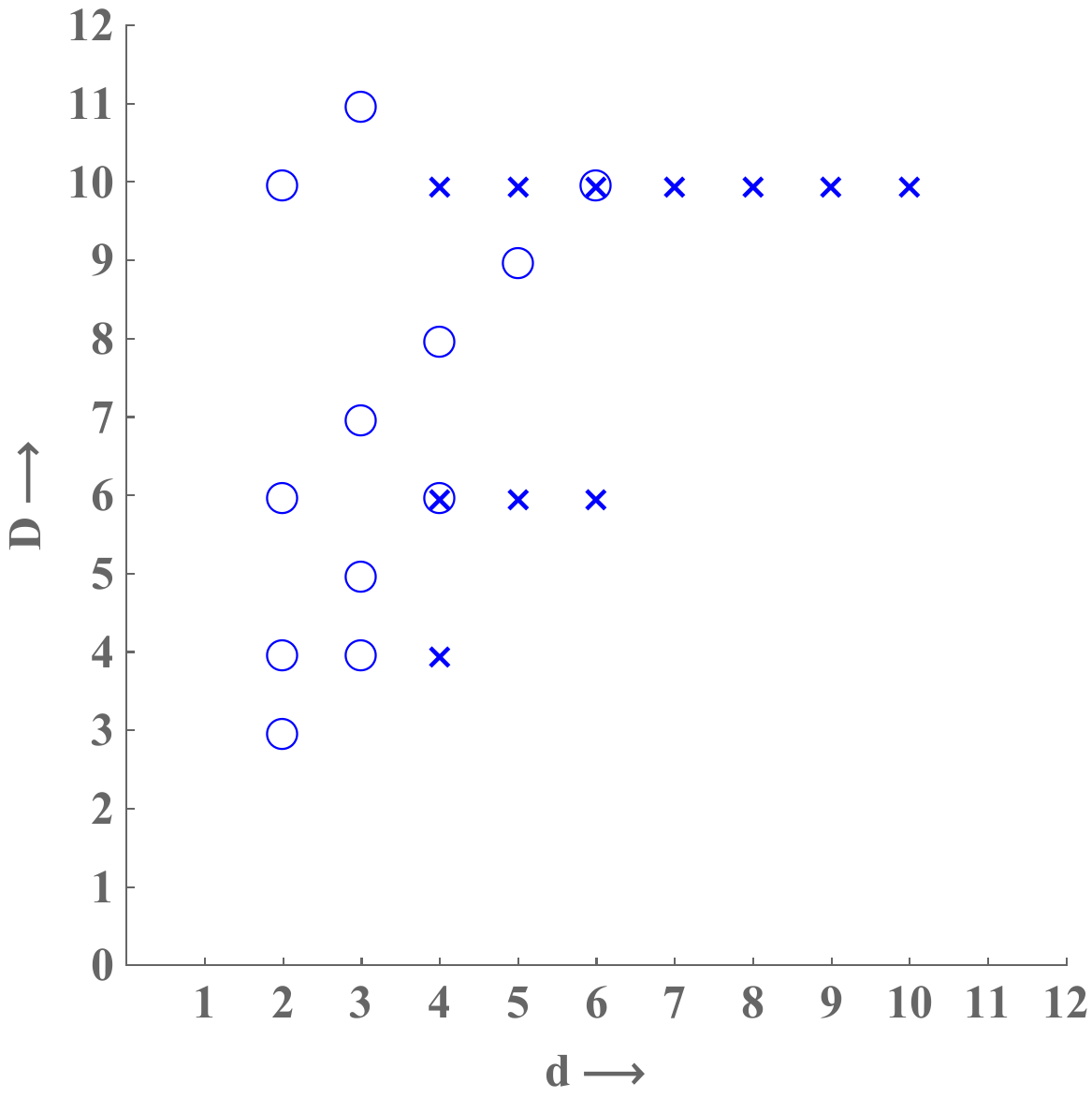}
\end{center}
\caption{The new brane scan.  All possible $d \ge 2$ scalar supermultiplets, denoted by circles, and vector supermultiplets, denoted by crosses, according to $D = d + $ the number of scalars.  }\label{newbs}
\end{figure}

Duff and myself not only found these 1/2 BPS p-branes but also classified them all \cite{Duff:1992hu, Duff:1994an}, which can be summarized in Figure \ref{newbs}.

Note that all these 1/2 BPS branes are found as solutions from the corresponding supergravities, the crosses in Figure \ref{newbs} represent the new 1/2 BPS branes not given in the previous old brane-scan in Figure \ref{oldbs}.

All these 1/2 BPS p-branes are the string solitons of the non-perturbative superstring theory and they are the corresponding non-perturbative superstring states. Just like the fundamental string or F-string, they are the basic dynamical objects of the complete non-perturbative string theory.  These objects including the strings are all intrinsically connected to each other. This discovery puts an end to the early assertion that strings have nothing to do with other branes.  In other words, if one wants to study strings, the other brane dynamics cannot be ignored in general and vice versa.

For Type II in D = 10, the newly discovered p-branes, except for the Type IIA NSNS 5-branes (also the D = 11 M5 brane) whose worldvolume modes give a tensor supermultiplet, are all vector supermultiplets, i.e., the supersymmetric Yang-Mills. These are nowadays called D-branes.

Polchinski et al \cite{Dai:1989ua} almost at the same time discovered also these branes  but via a completely different approach which was not widely accepted then.  That the branes found by the two different approaches were later recognized to be the same ones had to wait until the surge of the so-called second string revolution around 1995.  This was still due to the late Polchinski \cite{Polchinski:1995mt, Polchinski:1996nb}.

Polchinski et al \cite{Dai:1989ua}  applied then the T-duality, which was found from the  closed string theories, directly to open strings.  The consequence of this application requires the existence of D-branes. This at first  made people hardly acceptable since open strings cannot wind around compact directions and further the required hyperplanes identified as D-branes for consistency have no place in the perturbative string theories, giving the prejudice of then string community against the existence of higher dimensional branes. As such, the validity of this direct application of T-dualities to open strings was then questioned.

T-duality for closed strings in the simplest case can be understood in the following. Consider a closed string moving along a compactified circle with radius $R$ and the other one along a compactified circle with radius $\tilde R$, with the respective quantum momentum number $n, \tilde n $, and the respective string winding number around the circles as $w, \tilde w$.  If $R \tilde R =  \alpha'$ with the string slope parameter $\alpha'$ and we make the exchanges of $n \leftrightarrow \tilde w $ and
$w \leftrightarrow \tilde n$,  then these two string theories are either equivalent (as in the case of bosonic string theory) or one string theory is mapped to the other one (as in IIA and IIB).

From the view of string worldsheet, T -duality is nothing but the worldsheet electromagnetic or Hodge duality.  If we apply this to open string, it has the following consequence to its boundary conditions
\be
{\rm Neumann\,\, BC} \Longleftrightarrow {\rm Dirichlet \, \, BC}.
\ee

In other words, if we perform a T-duality along a direction for which the open string obeys initially the Neumann boundary condition, it will obey the Dirichlet one after this duality and vice versa. In other words, 
the two ends of an open string after certain number of such T-dualities will obey the Dirichlet boundary condition along these T-dual directions and the Neumann boundary condition along the rest directions. 
Therefore, the Dirichlet boundary conditions obeyed by the ends of the open string define the location of a hyperplane, on which the ends of the open string can move freely, along these Dirichlet directions.

The discovery of the string solitons associated with RR potentials not only validates  Polchinski et al's application of the T-duality found from closed strings to  open strings but also reveals indeed the existence of D-branes in string theories. 

The discovery of D-branes by Polchinski et al \cite{Dai:1989ua, Polchinski:1995mt} via open string T-duality, with respect to that as string solitons, has an advantage and usefulness in that in  weak string coupling, the open string so defined provides a perturbative description for the non-perturbative D-branes and this appears to be the first in the history of physics in that a perturbative description of the underlying non-perturbative objects can be provided in the region of small coupling.   We know that the massless modes of the open string is supersymmetric Yang-Mills and this is also consistent with what has been found as the zero modes of the solitonic D-branes which form the corresponding vector supermultilets.

The 1/2 BPS self-dual D3 brane along with its zero modes, namely the {\cal N = 4} super Yang-Mills in d = 4 found by Duff and myself \cite{Duff:1991pea} along with the non-abelian extension which can be obtained from the open string description \cite{Dai:1989ua, Polchinski:1995mt}, provides a basis for the AdS/CFT correspondence proposed later by Maldacena \cite{Maldacena:1997re}.

In summary, finding the stringy extended solitons, i.e.  the 1/2 BPS p-branes, from various supergravities, has the following significances:
1) establishing the intrinsic connection among these branes including the fundamental strings;
2) validating the use of the open string T-duality and as such leading to the useful description of the D-brane solitons in terms of perturbative open string when the string coupling is weak;
3) providing the basis for the AdS/CFT correspondence.

Moreover, this finding provides a basis for the various string dualities, which always exchange the fundamental string with its solitons,   and plays an important role for the existence of a unified theory called M-Theory \cite{Witten:1997fz}. 

By now, we hope that we have provided enough physical motivations to convince that finding the so-called 1/2 BPS basic extended objects associated with the various form potentials in
various supergravities is extremely important for understanding the non-perturbative properties of the underlying unified theory since they are the  objects which can be used to explore this unknown theory.

Without further ado, we will focus on finding these 1/2 BPS extended objects from various supergravities with maximal number of SUSY  in diverse dimensions in what follows. For those BPS extended solutions from supergravities with less number of SUSY and other aspects of these solutions in diverse dimensions, we refer to, for examples, \cite{Duff:1996hp, Youm:1997hw,Stelle:1998xg}. For black brane solutions and Non-BPS solutions\footnote{In general, the black solutions found are only good in the low energy limit, i.e., when the corresponding supergravity is valid. Unlike the SUSY preserving 1/2 BPS solutions, such kind of solutions valid in low energy limit does not necessarily imply their existence in the underlying non-perturbative UV complete theory. However, if the underlying theory is supersymmetric like supergravities,  the corresponding extremal or BPS solution usually preserves certain number unbroken supersymmetries and such a SUSY preserving solution, though the solution itself will be corrected when quantum and/or higher order corrections are included,  its existence as a state in the corresponding non-perturbative UV theory will in general be guaranteed.  In other words, only a SUSY preserving BPS solution can be a potential state in the underlying non-perturbative UV complete theory.  Precisely because of this, explicitly checking if a BPS solution preserves certain number of SUSY becomes important. },  we refer to \cite{Horowitz:1991cd} for black brane solutions in 10 dimensions, \cite{Duff:1993ye, Duff:1994an, Duff:1996hp, Youm:1997hw,Stelle:1998xg} for black p-brane solutions in diverse dimensions and \cite{Lu:2004ms} for non-supersymmetric p-brane solutions including non-BPS ones in diverse dimensions.

\section{The brane  sigma-model action}

Before we move to find the 1/2 BPS p-branes, we would like to give a brief introduction in the following to the bosonic part of the supersymmetric p-brane sigma-model action for those branes in the old brane-scan.  This bosonic part will be also good for finding brane solutions  from supergravities for the branes in the new brane scan since so long the static SUSY preserving 1/2 BPS p-brane is concerned, the other worldvolume fields such as the possible vector or tensor along with the fermionic ones are not exited, therefore setting to vanish.

Just like a point particle moving in spacetime gives a worldline, a string moving in spacetime gives a (1 + 1)-dimensional worldsheet,  a general p-brane moving in spacetime gives a (1 + p)-dimensional worldvolume.

 \begin{figure}[t]
\begin{center}
\includegraphics[scale= 0.4]{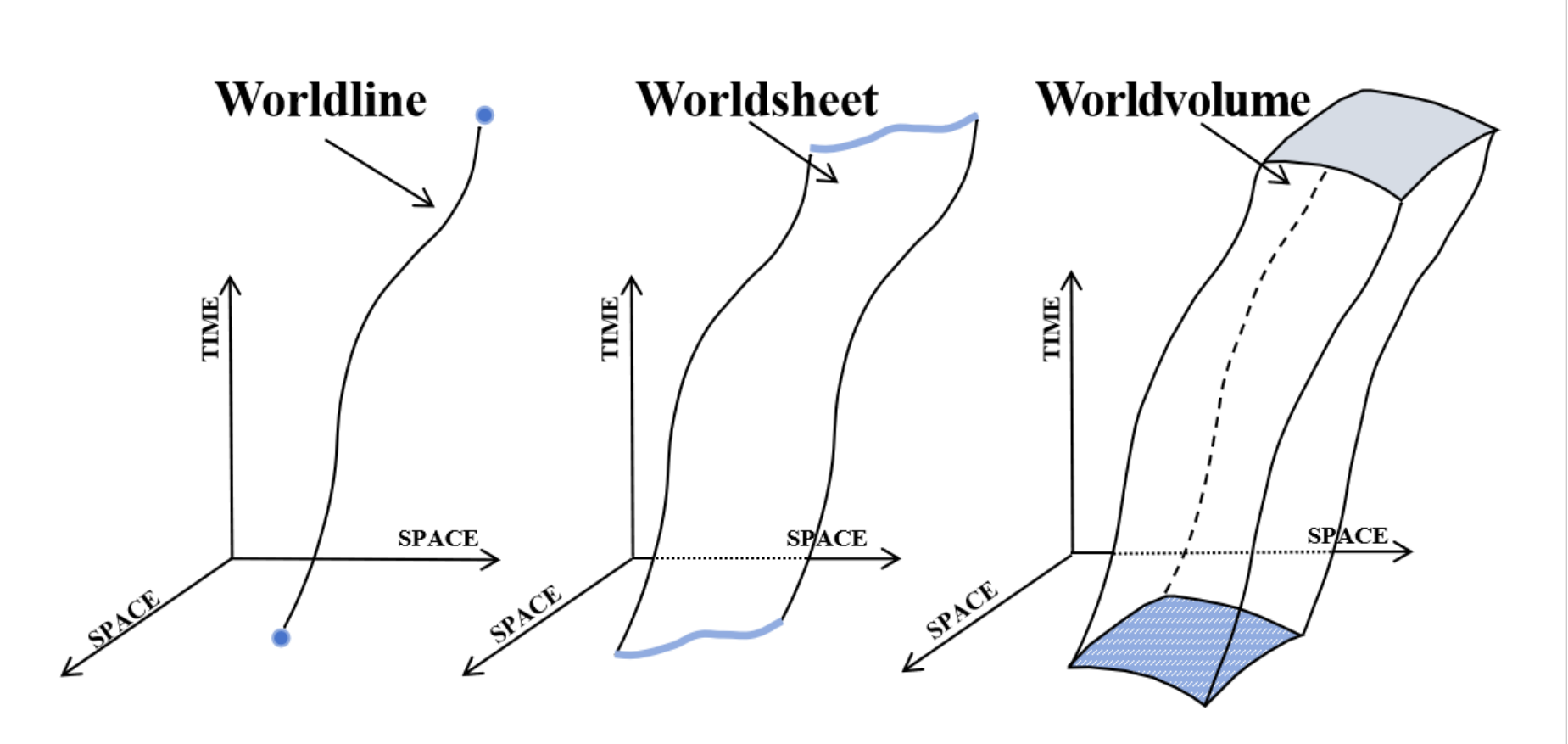}
\end{center}
\caption{Particles, strings and branes  }\label{worldv}
\end{figure}

For those super p-branes in the old brane-scan, the worldvolume action of a super p-brane moving in a curved superspace with a superspace coordinate $Z^{M} = (x^{\mu}, \theta^{\alpha})$ is \cite{Bergshoeff:1987cm, Achucarro:1987nc}
\bea\label{sp-action}
S_{d} &=& T_{d} \int d^{d} \sigma \left( -\frac{1}{2}  \sqrt{ - \gamma} \gamma^{ij} E_{i}\,^{a} E_{j}\,^{b} \eta_{ab} + \frac{d - 2}{2} \sqrt{ - \gamma} \right.\nn
&\,& \left. \qquad + \frac{1}{d !} \epsilon^{i_{1} \cdots i_{d}} E_{i_{1}}\,^{A_{1}} \cdots E_{i_{d}}\,^{A_{d}} C_{A_{1} \cdots A_{d}} \right),
\eea
where the first term is the usual kinetic or the volume term determined by the metric, the second the cosmology one and the third the Wess-Zumino or the volume one determined by the total antisymmetric density $\epsilon^{i_{1} \cdots i_{d}}$.

In the above, $E_{M}\,^{A}$ is the supervielbein  with the superspace world indices $M = \mu, \alpha$ and the tangent space indices $A = a, \alpha$.  We also define the worldvolume pull-back as $E_{i}\,^{A} = \partial_{i} Z^{M} E_{M}\,^{A}$ with $i, j = 0, 1, \cdots p$ the worldvolume indices. Note here $d = 1 + p$, $\mu =0, 1 \cdots D - 1$ with $D$ the spacetime dimension and $\alpha$ are the spinor indices of spacetime spinor coordinate $\theta$. $C_{A_{1} \cdots A_{d}} (Z)$ is the super d-form potential.

The target-space symmetries of this action are super diffeomorphisms, Lorentz invariance and d-form gauge invariance. The worldvolume symmetries are ordinary diffeomorphisms and the $\kappa$-symmetry defined as
\be
\delta Z^{M} E_{M}\,^{a} = 0, \quad \delta Z^{M} E_{M}\,^{\alpha} =  (1 + \Gamma)^{\alpha}\,_{\beta} \kappa^{\beta},
\ee
with
\be
\Gamma^{\alpha}\,_{\beta} = \frac{( - )^{d (d - 3)/4}}{d! \sqrt{- \gamma}} \epsilon^{i_{1} \cdots i_{d}} E_{i_{1}}\,^{a_{1}} \cdots E_{i_{d}}\,^{a_{d}} \left(\Gamma_{a_{1} \cdots a_{d}}\right)^{\alpha}\,_{\beta}.
\ee
\medskip
If we focus on the bosonic part of the above action (\ref{sp-action}), it gives
\bea\label{bpa}
S_{d} &=& T_{d} \int d^{d} \sigma \left( -\frac{1}{2}  \sqrt{ - \gamma} \gamma^{ij} \partial_{i} X^{\mu} \partial_{j} X^{\nu} G_{\mu\nu} (X) + \frac{d - 2}{2} \sqrt{ - \gamma} \right.\nn
&\,& \left. \qquad + \frac{1}{d !} \epsilon^{i_{1} \cdots i_{d}} \partial_{i_{1}} X^{\mu_{1}} \cdots \partial_{i_{d}} X^{\mu_{d}} C_{\mu_{1} \cdots \mu_{d}} \right),
\eea
where $G_{\mu\nu}$ is the background metric in the so-called p-brane frame (its relation with the Einstein frame metric will be given latter on) and $C_{\mu_{1} \cdot \mu_{d}}$ the d-form potential.

The above action reminds that a p-dimensional object, when it carries a U(1) charge, must couple with a (1 + p)-form potential when the U(1) local symmetry is insisted.  It is clear that the action (\ref{bpa}) is invariant up to a surface term (the EOM is invariant) under the gauge transformation
\be
C_{d} \to C'_{d} = C_{d} + d \Lambda_{d - 1},
\ee
where the (d - 1)-form $\Lambda_{d - 1}$ is the gauge transformation parameter.

This is just like in QED. A local U(1) symmetry must imply that a U(1) charged particle (or object) couples with the U(1) gauge potential (or higher form gauge potential) for consistency.

Noether theorem says that a global U(1) symmetry gives a conserved charge. If this symmetry can be promoted to a local one, i.e., a U(1) gauge symmetry, there must exist a U(1) form gauge potential associated with the corresponding conserved current for consistency and their interaction can also be easily determined via the standard current and gauge potential coupling or the minimal coupling.

Let us see how the Wess-Zumino action is obtained from this coupling. For a point charge $q$ moving in bulk spacetime along its worldline $X^{\mu} (\tau)$, the current produced by this charge at a spacetime point $x$ is
\be
j^{\mu} (x) = q \int d\tau ~\partial_{\tau} X^{\mu} (\tau) ~\delta^{(D)} \left(x - X (\tau)\right),
\ee
then the coupling is given by $j^{\mu} (x) A_{\mu} (x)$ and the Wess-Zumino action is given as
\bea
S_{WZ} &=& \int d^{D} x~ j^{\mu} (x) A_{\mu} (x)\nn
&=& q \int d\tau ~\partial_{\tau} X^{\mu} (\tau) A_{\mu} (X).
\eea

For a string with its line charge density $\mu_{1}$ moving in bulk spacetime with its worldsheet $X^{\mu} (\tau, \sigma)$, the current produced at a given spacetime point $x$ is
\be
j^{\mu\nu} (x) = \mu_{1} \int d^{2}\sigma ~\epsilon^{ij} \partial_{i} X^{\mu} \partial_{j} X^{\nu} \delta^{(D)} \left(x - X(\tau, \sigma)\right),
\ee

The coupling is then $j^{\mu\nu} (x) B_{\mu\nu} (x)/2!$ and the Wess-Zumino action is
\bea
S_{WZ} &=& \int d^{D} x ~j^{\mu\nu} (x) B_{\mu\nu} (x)\nn
&=& \frac{\mu_{1}}{2!} \int  d^{2} \sigma~ \epsilon^{ij} \partial_{i} X^{\mu} \partial_{j} X^{\nu} B_{\mu\nu} (X).
\eea

In general, for a p-brane with its p-volume charge density $\mu_{p}$ moving in bulk spacetime, the current produced at point $x$ is
\be\label{p-brane-current}
j^{\mu_{1}\cdots \mu_{p + 1}} (x) = \mu_{p} \int d^{p + 1} \sigma~ \epsilon^{i_{1} \cdots i_{p + 1}} \partial_{i_{1}} X^{\mu_{1}} \cdots \partial_{i_{p + 1}} X^{\mu_{p + 1}} \delta^{(D)} \left(x - X(\sigma)\right).
\ee
this gives the coupling
\be
\frac{1}{(p + 1)!} j^{\mu_{1} \cdots \mu_{p + 1}} (x) C_{\mu_{1} \cdots \mu_{p + 1}},
\ee
and the Wess-Zumino action is

\bea
S_{WS} &=& \frac{1} {(p + 1)!} \int d^{D} x~ j^{\mu_{1} \cdots \mu_{p + 1}} (x) C_{\mu_{1} \cdots \mu_{p + 1}}\nn
&=& \frac{\mu_{p}}{(p + 1)!} \int d^{p + 1} \sigma ~\epsilon^{i_{1} \cdots i_{p + 1}} \partial_{i_{1}} X^{\mu_{1}} \cdots \partial_{i_{p + 1}} X^{\mu_{p + 1}} C_{\mu_{1} \cdots \mu_{p + 1}} (X).\nn
\eea
and the conserved form charge is
\bea
Z^{\mu_{1} \cdots \mu_{p}} &\equiv& \int d^{D -1} x\, j^{0\, \mu_{1} \cdots \mu_{p}} (x) \nn
&=& \mu_{p} \int d^{p}\sigma\, \epsilon^{0 \,i_{1} \cdots i_{p}} \partial_{i_{1}} X^{\mu_{1}} \cdots \partial_{i_{p}} X^{\mu_{p}},
\eea
where we have taken  $\sigma^{0} = x^{0}$.
For example, if we take the source as a static one, say, along $X^{i} = \sigma^{i}$, we have
\be
Z^{12\cdots p} = \mu_{p}  V_{p},
\ee
where $V_{p}$ is the spatial p-volume of the brane.

\section{1/2 BPS p-branes from various supergravities in diverse dimensions}

Before we begin this section, a few remarks follow.  Firstly, without the understanding and the physics guidance given in Section 1,  we don't have the physical motivation to seek the 1/2 BPS p-brane solutions from various supergravities and to find their connection, which was then unpopular, to the fundamental strings.  Note also that the supergravities were found around 1970s', if the purpose is merely for finding solutions, this would have had been done long before around the end of 1980s'. Even purely for the purpose of finding stable BPS solutions, if there are no the physical guidance and the clear physical picture, finding such solutions would hardly be possible and at least it should be an extremely difficult task given the higher non-linearity and the complexity of supergravity theories. For example, the Lagrangian for the 10 dimensional IIA supergravity as given in \cite{Huq:1983im}, not mentioning the SUSY transformations for various fields involved, is much more complicated than the usual Einstein gravity.

\medskip
\noindent
{\bf The generality}

\medskip

Let us discuss some general features expected for the p-brane solutions from the supergravities in diverse dimensions.

Suppose that we begin with an empty D-dimensional Minkowski spacetime.  In other words, we have the D-dimensional Poincare symmetry $P_{D}$. Now consider to place a p-brane source ($p < D - 1$) in this spacetime.

Due to its tension, therefore its mass, and its charge, this brane will curve the spacetime and give rise to a  (p + 1)-form potential or (p + 2)-form field strength around it. 

\begin{figure}[t]
\begin{center}
\includegraphics[scale= 0.8]{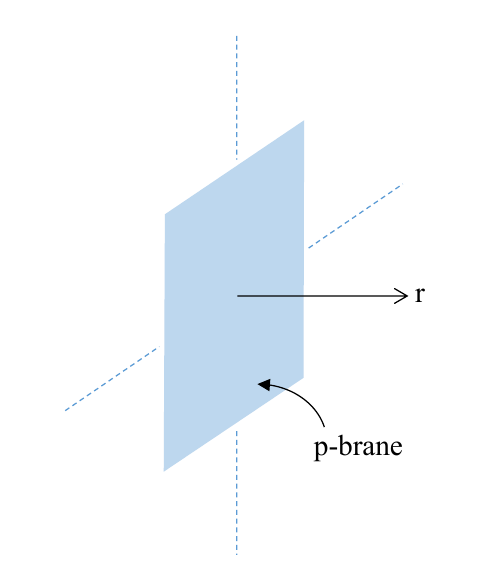}
\end{center}
\caption{1/2 BPS p-brane configuration}\label{p-brane}
\end{figure}

We are seeking a static and stable BPS configuration and this must require that the brane be infinitely extended along its p-spatial directions and the brane tension be equal to its charge density in certain units such that the attraction due to its tension can cancel the repulsion due to its charge.  Otherwise, it is impossible to balance the attraction against the repulsion and to give rise to a stable configuration. Picture-wise, this p-brane configuration is represented in Figure \ref{p-brane}.

Given what has been said, for a coupled system of a p-brane and the background fields involving gravity, finding a static SUSY preserving 1/2 BPS p-brane vacuum-like configuration is not so obvious at a first look partly because of the higher non-linearity of the system (unlike the case finding the electric field of a given charge from the linear Maxwell equations in 4 dimensions). Nevertheless, the physical ground as we gave earlier implies the existence of such a SUSY preserving configuration. 
Once such a brane source is placed in spacetime, we expect the original underlying symmetry $P_{D}$ to be broken to $P_{d} \times SO(D - d)$ with $d = 1 + p$ and $P_{d}$ being the d-dimensional Poincare group. If we split the D-coordinates as $x^{M} = (x^{\mu}, x^{m})$ with $\mu =0, 1, \cdots p$ and $m = d, d + 1, \cdots D - 1$. We have therefore the most general ansatz for the D-dimensional Einstein frame metric, respecting this residue symmetry, to be
\be\label{pmetric}
ds^{2} = e^{2 A(r)} \eta_{\mu\nu} d x^{\mu} d x^{\nu} +  e^{2 B(r)} \delta_{mn} d x^{m} d x^{n},
\ee
with $r = \sqrt{\delta_{mn} x^{m} x^{n}}$ the radial coordinate along the transverse directions of the brane.

The ansatz for the dilaton $\phi$ is $\phi = \phi (r)$. If the brane is treated as an electric-like source, then the ansatz for the (1 + p)-form potential is
\be\label{d-formp}
A_{01\cdots p} = - \left(e^{C(r)} - 1\right).
\ee
As described above, we are looking for the static field configuration produced by the brane source which is also static and the whole system preserves certain number of SUSY (here actually 1/2). 
For this reason, the worldvolume fields of the brane are all frozen except for the embedding coordinates describing the location of the brane. 
For the infinitely extended brane, we expect to have
\be
\sigma^{\mu} = X^{\mu}, \qquad  X^{m} = 0,
\ee
i.e., the brane is located at $r = 0$ along the directions transverse to the brane.

Since the brane configuration along with the source is invariant under $P_{d} \times SO(D - d)$ and this is a vacuum-like configuration,  the spacetime (bulk) fermionic fields and the worldvolume fermionic fields must be both set to vanish. In other words, only the relevant bosonic fields, which remain invariant under the residue symmetry $P_{d} \times SO(D - d)$, are relevant to this configuration.

This configuration is expected to preserve some SUSY, therefore the transformations of both bosonic fields and fermionic fields are also expected to vanish such that this configuration remains invariant under the unbroken SUSY.

As will be seen, this remains true automatically for the bosonic fields once the fermionic ones are set to vanish and the requirement  that the transformed fermionic fields remain so determines how many SUSY are preserved for this configuration. 

Given what has been said, we first find p-brane solutions in diverse dimensions and then discuss some specific cases for illustrating the number of the SUSY preserved \cite{Duff:1993ye, Duff:1994an}.

\medskip
\noindent
{\bf p-branes in diverse dimensions}
\medskip

Given that the fermionic fields for both bulk and worldvolume are set to vanish, we need only to consider the relevant  bosonic fields\footnote{The only relevant bulk fields for a simple p-brane are the metric and dilaton due to the brane tension as well as charge which give rise to the gravity multiplet which includes the metric and dilaton in the bulk, and the $(p + 1)$-form potential which couples with this brane source due to its charge. All other fields are irrelevant to this simple brane source and are set to vanish.  One has to have these in minds, otherwise one doesn't know how to begin with given the complication of supergravity. }  in the corresponding supergravity action plus the bosonic action of brane. In other words, we need to consider only bosonic action of the combined bulk and brane, i.e., $I_D (d) + S_d$, where the Einstein-frame bulk action is
\be \label{bulkA}
I_D (d) = \frac{1}{2 \kappa^2} \int d^D x \sqrt{- g}\left( R - \frac{1}{2} (\partial \bar\phi)^2 - \frac{1}{2 ( d + 1)!} e^{- \alpha (d) \bar\phi} F_{d + 1}^2 \right),\ee
and the brane sigma-model one is
\bea \label{p-braneA} S_d &=& T_p \int d^d \sigma  \left[ - \frac{1}{2} \sqrt{- h}\, h^{\alpha\beta} \partial_\alpha X^M \partial_\beta X^N  G_{MN}
+ \frac{d - 2}{2} \sqrt{- h}\right.
 \nn &\,& \qquad \qquad \quad - \left.\frac{1}{d! } \epsilon^{\alpha_1 \cdots \alpha_d} \partial_{\alpha_1} X^{M_1} \cdots \partial_{\alpha_d} X^{M_d} A_{M_1 \cdots M_d}\right].
\eea
In the above, both the p-brane $\sigma$-model metric $G_{MN}$ and the Einstein metric $g_{MN}$ are asymptotically flat and the possible vacuum expectation value of $\phi$ or its asymptotically one is absorbed into the respective factors.  So the $\kappa$ and the brane tension $T_{p}$ are both physical.

In the above, $F_{d + 1} = d A_d,  d = p + 1$ and the p-brane frame metric $G_{MN} = g_{MN} e^{\alpha (d) \bar\phi/d}$ with $g_{MN}$ the Einstein frame metric (See \cite{Duff:1994an} for deriving this relation) and $\bar \phi = \phi - \phi_{0}$ with $\langle \phi \rangle = \phi_{0}$ and the string coupling $g_{s} = e^{\phi_{0}}$. We have also the following: For D = 11, $\alpha (d) = 0$; For D = 10, if we choose to relate the relevant parameters to string ones, we have $ 2\kappa^{2} = (2 \pi)^{7} \alpha'^{4} g^{2}_{s}$ and
\bea
\alpha (d) &=& \frac{3 - p}{2}\,  ({\rm NS(NS)}\, {\rm p\, brane}),\quad \alpha (d) = \frac{p - 3}{2}\,  ( {\rm Dp\, brane}),\nn
T_{\rm F} &=& \frac{1}{2\pi \alpha'}, \quad T_{\rm NS 5} = \frac{1} {g^{2}_{s} (2\pi)^{5} \alpha'^{3}},\quad T_{\rm Dp} = \frac{1}{g_s (2\pi)^p \alpha'^{(1 + p)/2}}.
\eea
 For a 1/2 BPS p-brane in diverse dimensions with the corresponding supergravity having a maximal SUSY, we have in general\footnote{For supergravities with less number of SUSY, see \cite{Lu:1995cs, Lu:1995yn}.}
 \be\label{alpha-p}
\alpha^{2} (d) = 4 - \frac{2 d \tilde d}{D - 2},
\ee
with $\tilde d = D - 2 - d$.

From the action $I_{D} (d) + S_{d}$ with $I_{D} (d)$ given in (\ref{bulkA}) and $S_{d}$ given in (\ref{p-braneA}),   the equations of motion (EOM) are,  for the metric
 \bea\label{Enstein}
  &&R_{MN} - \frac{1}{2} g_{MN} R - \frac{1}{2}\left(\partial_{M} \bar\phi \partial_{N} \bar\phi - \frac{1}{2} g_{MN} (\partial\bar \phi)^2\right)\nn
 && - \frac{1}{2 d!}\left(F_{M M_1 \cdots M_d} F_{N}\,^{M_1 \cdots M_d} - \frac{1}{2 (d + 1)} g_{MN} F^2_{d + 1} \right) e^{ - \alpha (d)\bar \phi} 
 = \kappa^2 T_{MN} ({\rm p\,brane}),
 \eea
 for the dilaton
 \bea\label{dilaton-EOM} &&\partial_M (\sqrt{- g} g^{MN} \partial_N \bar\phi) + \frac{\alpha (d)}{2 (d + 1)!} \sqrt{- g} e^{- \alpha (d)\bar \phi } F^2_{d + 1}\nn
 &&= \frac{\alpha (d) \kappa^2 T_p}{d} \int d^d \sigma \sqrt{- h} \, h^{\alpha\beta} \partial_\alpha X^M \partial_\beta X^N g_{MN} e^{\alpha(d)\bar\phi/d} \delta^{(D)} (x - X),
 \eea
 and for the d-form potential
 \bea\label{d-formEOM} &&\partial_M \left(\sqrt{- g} e^{-\alpha (d)\bar\phi} F^{M M_1 \cdots M_d} \right) \nn
 && = 2 \kappa^2 T_p \int d^d \sigma \epsilon^{\alpha_1 \cdots \alpha_d} \partial_{\alpha_1} X^{M_1} \cdots \partial_{\alpha_d} X^{M_d} \delta^{(D)} (x - X).\eea
 In the above, the energy-momentum tensor with up indices is
 \be \label{emtensor}
 T^{MN} ({\rm p\,brane}) = - T_p \int d^d \sigma \sqrt{- h}\, h^{\alpha\beta} \partial_\alpha X^M \partial_\beta X^N e^{\alpha (d)\bar\phi/d} \frac{\delta^{(D)} (x - X)}{\sqrt{- g}}.
 \ee
 The p-brane EOM are, for $X^M$
 \bea\label{p-braneEOM} 
 &&\partial_\alpha \left(\sqrt{-h} \, h^{\alpha\beta} \partial_\beta X^N g_{MN} e^{\alpha (d) \bar\phi/d}\right)
  -\frac{1}{2}\sqrt{-h}\, h^{\alpha\beta} \partial_\alpha X^N \partial_\beta X^P \partial_M \left(g_{NP} e^{\alpha (d) \bar\phi/d} \right)\nn
&&  - \frac{1}{d!} \epsilon^{\alpha_1 \cdots \alpha_d} \partial_{\alpha_1} X^{M_1} \cdots \partial_{\alpha_d} X^{M_d} F_{M M_1 \cdots M_d} = 0,\eea
and for the worldvolume metric $h_{\alpha\beta}$
 \be h_{\alpha\beta} = \partial_\alpha X^M \partial_\beta X^N g_{MN} e^{\alpha (d) \bar\phi/d}.\ee
 For a static p-brane ($X^{\mu} = \sigma^{\mu}, X^{m} = 0$), we have
 \be
 h_{\mu\nu} = g_{\mu\nu} e^{\alpha(d) \bar\phi/d} =  \eta_{\mu\nu} e^{2 A + \alpha (d) \bar\phi/d},
 \ee
 where $\mu, \nu = 0, 1, \cdots p$ and the metric $g_{MN}$ is given by (\ref{pmetric}).   So
 \be
 h = \det h_{\mu\nu} = - e^{2 A d + \alpha (d) \bar\phi},\quad g = \det g_{MN} = - e^{2 A d + 2 B (D - d)}.
 \ee
 The energy-momentum tensor (\ref{emtensor}) is now
 \bea
 T_{\mu\nu} &=& - T_{p} \,\eta_{\mu\nu} \, e^{2 A - B (D- d) + \alpha(d) \bar\phi/2}\, \delta^{(D - d)} (x_\bot ),\nn
 T_{mn} &=& 0.
 \eea
 From (\ref{d-formp}), we have
 \bea
 F_{\mu M_{1} \cdots M_{d}} F_{\nu}\,^{M_{1} \cdots M_{d}} &=& - d!\,\eta_{\mu\nu} e^{- 2 A(d - 1) - 2 B + 2 C} \delta^{mn} \partial_{m} C \partial_{n} C,\nn
 &\,&\nn
 F_{m M_{1} \cdots M_{d}} F_{n}\,^{M_{1} \cdots M_{d}} &=& - d! \, e^{- 2 A d + 2 C}\, \partial_{m} C \partial_{n} C ,\nn
 &\,& \nn
 F^{2}_{d + 1} &=& - (d + 1)!\, e^{- 2 B - 2 A d + 2 C}\, \delta^{mn} \partial_{m} C \partial_{n} C .\nn
 \eea
With the given ansatz, the $\mu\nu$-components of Einstein EOM (\ref{Enstein}) are
\bea\label{EOM1}
&& \delta^{mn} \left[(d - 1) \partial_m \partial_n A + (\tilde d + 1) \partial_m \partial_n B + \tilde d (d - 1) \partial_m A \partial_n B
 \right.\qquad  \nn
&& + \frac{d (d - 1)}{2} \partial_m A \partial_n A + \frac{\tilde d (\tilde d + 1)}{2} \partial_m B \partial_n  B + \frac{1}{4} \partial_m \bar\phi \partial_n \bar\phi
   \nn
 &&\left. + \frac{1}{4} e^{- 2 d A + 2 C - \alpha (d) \bar\phi} \partial_m C \partial_n C \right]  = - \kappa^2 T_p e^{ - B \tilde d + \alpha (d)\bar\phi/2} \delta^{(D - d)} (x_{\bot}),
 \eea
and the $mn$-components
\bea\label{EOM2} &&-\tilde d\left(\partial_m \partial_n B -   \delta_{mn} \delta^{kl} \partial_k \partial_l B \right) - d \left(\partial_m \partial_n A - \delta_{mn} \delta^{kl} \partial_k \partial_l A\right) \nn
&&- d\left(\partial_m A \partial_n A - \frac{d + 1}{2} \delta_{mn} \delta^{kl} \partial_k A \partial_l A\right) + \tilde d \left(\partial_{m} B \partial_{n} B + \frac{\tilde d - 1}{2} \delta_{mn} \delta^{kl} \partial_{k} B \partial_{l} B \right)\nn
 && + d \left(\partial_m A \partial_n B + \partial_m B \partial_n A + (\tilde d - 1) \delta_{mn} \delta^{kl} \partial_k A \partial_l B \right) - \frac{1}{2} \partial_m\bar\phi \partial_n \bar\phi + \frac{1}{4} \delta_{mn} \delta^{kl} \partial_k\bar \phi \partial_l \bar\phi \nn
&&- \frac{1}{2} e^{- 2 d A  + 2 C - \alpha(d) \bar\phi} \left[- \partial_m C \partial_n C + \frac{1}{2} \delta_{mn} \delta^{kl} \partial_k C \partial_l C\right] = 0,
\eea
which can be rewritten as
\bea\label{EOM2-0}
&& - \partial_{m} \partial_{n} (d A + \tilde d B) + \partial_{m} (d A + \tilde d B) \partial_{n} B + d \partial_{m} (B - A) \partial_{n} A  - \frac{1}{2} \partial_{m}\bar\phi\partial_{n}\bar\phi \nn
&& + \frac{1}{2} e^{- 2 d A 
+ 2 C - \alpha \bar\phi} \partial_{m} C\partial_{n} C\nn
&& + \delta_{mn} \delta^{kl} \left[\partial_{k} \partial_{l} (d A + \tilde d B) + d (\tilde d - 1) \partial_{k} A \partial_{l} B + \frac{d (d + 1)}{2} \partial_{k} A \partial_{l} A + \frac{\tilde d (\tilde d - 1)}{2} \partial_{k} B  \partial_{l} B \right.\nn
&& \left.+ \frac{1}{4} \partial_{k} \bar\phi \partial_{l} \bar \phi - \frac{1}{4} e^{- 2 d A + 2 C - \alpha \bar\phi} \partial_{k} C \partial_{l} C\right] = 0.
\eea
In deriving the above two equations, given the metric form,
\be d s^2 = e^{2 A (r)} \eta_{\mu\nu} d x^\mu d x^\nu + e^{2 B (r)} \delta_{mn} d y^m d y^n,\ee
with $\mu, \nu = 0, 1, \cdots d - 1$ and $m, n = d, \cdots D - 1$, we have used the following
\bea \label{ricci-tensor}
R_{\mu\nu} &=& -\eta_{\mu\nu}  e^{2 (A - B)} \delta^{mn}\left(\partial_m \partial_n A + d \,\partial_m A \partial_n A + \tilde d\, \partial_m A \partial_n B\right),\\
 R_{mn} &=&  - \tilde d\, \partial_m \partial_n B - \delta_{mn} \delta^{kl} \partial_k\partial_l B - d\, \partial_m \partial_n A - d\, \partial_m A \partial_n A\nn
 && + d \left(\partial_m A \partial_n B + \partial_m B \partial_n A - \delta_{mn} \delta^{kl} \partial_k A \partial_l B\right)
 + \tilde d \left(\partial_m B \partial_n B - \delta_{mn} \delta^{kl} \partial_k B \partial_l B\right), \nn
 &&\\
 R &= &- e^{- 2 B}\delta^{kl} \left[2~ d ~\partial_k\partial_l A + d (d + 1) ~\partial_k A \partial_l A + 2 ~d ~\tilde d ~\partial_k A \partial_l B   + 2~  (\tilde d + 1)~\partial_k \partial_l B     \right.\nn
       &\,&\qquad  \qquad\quad  \left.  + \tilde d~ (\tilde d + 1) \partial_k B \partial_l B\right].
  \eea
In the above $\tilde d = D - d - 2$.

The EOM for the dilaton (\ref{dilaton-EOM}) reduces to
\bea\label{EOM3}
&&\delta^{mn} \partial_{m} \left(e^{A d + B \tilde d} \partial_{n}\bar\phi\right) - \frac{ \alpha(d)}{2} e^{ B \tilde d -  A d - \alpha(d) \bar\phi + 2 C} \delta^{mn} \partial_{m} C \partial_{n} C\nn
&&= \alpha (d) ~\kappa^{2} ~T_{p}~ e^{A d + \alpha (d) \bar\phi/2} \delta^{(D - d)} (x_{\bot}),
\eea
and the EOM for the d-form potential (\ref{d-formEOM}) reduces to
\be\label{EOM4}
\delta^{mn} \partial_{m}\left(e^{B \tilde d -  A d - \alpha (d) \bar\phi + 2 C} \partial_{n} e^{- C}\right) =  - 2~ \kappa^{2}~ T_{p}~ \delta^{(D - d)} (x_{\bot}),
\ee
and the EOM for the p-brane (\ref{p-braneEOM}) becomes now
\be
\partial_{m} \left(e^{C} - e^{Ad + \alpha(d) \bar\phi/2}\right) = 0,
\ee
which is usually called ``no force'' condition.

Note that $A (r \to \infty) = B(r \to \infty) = \bar \phi(r \to \infty) = 0$.  If we choose $C (r \to \infty) = 0$ [a proper choice for which $A_{01\cdots p} (r \to \infty) \to 0$], then from the last equation, we have
\be\label{CAP}
C = Ad + \alpha (d) \bar\phi/2.
\ee
With this, we have from (\ref{EOM4})
\be\label{EOM4-1}
\delta^{mn} \partial_{m} \left(e^{d A  + \tilde d B} \partial_{n} e^{- C} \right) = - 2 \kappa^{2} T_{p} \, \delta^{(D - d)} (x_{\bot}).
\ee
From (\ref{EOM3}) and combined with the above, we have
\be
\delta^{mn} \partial_{m}\left[e^{d A + \tilde d B} \partial_{n} \left(C - \frac{2 \bar\phi}{\alpha (d)}\right)\right] = 0,
\ee
or
\be\label{EOM-A}
\delta^{mn} \partial_{m}\left[e^{d A + \tilde d B} \partial_{n} \left(A - \frac{\tilde d}{2 (d + \tilde d)} C \right)\right] = 0,
\ee
where we have used (\ref{CAP}) and (\ref{alpha-p}). Note that here $A, B, C, \phi$ are the functions of variable $r$ only, so we have
\bea\label{formula1}
\delta^{mn} \partial_{m} \left( e^{d A + \tilde d B} \partial_{n} F (r)\right) &=& e^{d A + \tilde B} \left(F'' + \frac{\tilde d + 1}{r} F' + (d A + \tilde d B)' F'\right),\nn
\delta^{mn} \partial_{m} \partial_{n} F & =& F'' + \frac{\tilde d + 1}{r} F',
\eea
where $F' \equiv d F/d r, F'' \equiv d ^{2} F/d r^{2}$. Then the EOM (\ref{EOM1}), using (\ref{EOM4-1}) to replace the $\delta$-function on the right, can be rewritten as
\bea
0 &= &  (d A + \tilde d B)'' + \frac{\tilde d + 1}{r} (d A + \tilde d B)' + \frac{1}{2} (d A + \tilde d B)'^{2}  - \frac{1}{2} A' (d A + \tilde d B)' \nn
&& - (A - B)'' - \frac{\tilde d + 1}{r} (A - B)' - \frac{\tilde d}{2} (A - B)' B' +  \frac{1}{2} \left(C'' + \frac{\tilde d + 1}{r} C' \right) \nn
&&- \frac{d C'}{\alpha^{2}} \left(A - \frac{\tilde d}{2 (d + \tilde d)} C\right)' + \frac{d A'}{\alpha^{2}} (d A - C)' + \frac{1}{2} \left( d A + \tilde d B\right)' C',
\eea
where we have used (\ref{CAP}) to replace $\bar\phi$ by $A$ and $C$.  This can be further rewritten as
\bea\label{EOM1-1}
&& \left(B - A + \frac{C}{2}\right)'' + \frac{\tilde d + 1}{r} \left(B - A + \frac{C}{2}\right)' + \frac{\tilde d}{2} B' \left(B - A + \frac{C}{2}\right)'\nn
&&+ (d A + \tilde d B)'' + \frac{\tilde d + 1}{r} (d A + \tilde d B)' + \frac{1}{2} (d A + \tilde d B)'^{2}  - \frac{1}{2} \left( A - \frac{C}{2} \right)'  (d A + \tilde d B)'\nn
&&+ \frac{d }{\alpha^{2}} (d A - C)' \left(A- \frac{\tilde d}{2 (d + \tilde d)} C\right)' = 0.
\eea
For the $m\neq n$ components of (\ref{EOM2-0}), we have
\bea\label{EOM2-1}
&&(d A + \tilde d B)'' - \frac{1}{r} (d A + \tilde d B)'   - B' \left(d A  + \tilde d B\right)'  - d A' \left(B -  A + \frac{C}{2}\right)' \nn
&& + \frac{2 d}{\alpha^{2}} \left( d A - C\right)' \left(A - \frac{\tilde d}{2 (d + \tilde d)} C\right)'= 0,
\eea
where we also use (\ref{CAP}) to replace $\bar\phi$ in terms of $A$ and $C$.  The $m = n$ components of (\ref{EOM2-0}) give
\bea\label{EOM2-2}
&&(d A + \tilde d B)'' + \frac{\tilde d + 1}{r} (d A + \tilde d B)' + \frac{1}{2} (d A + \tilde d B)'^{2} - \frac{1}{2} B' (d A + \tilde d B)' \nn
&& - \frac{d}{2} A' \left(B - A + \frac{C}{2}\right)' + \frac{d}{\alpha^{2}} (d A - C)' \left(A - \frac{\tilde d}{2 (d + \tilde d)} C\right)' = 0.
\eea
We subtract (\ref{EOM2-2}) from (\ref{EOM1-1}) to give
\be\label{EOM1-1-1}
\left(B - A + \frac{C}{2}\right)'' + \frac{\tilde d + 1}{r} \left(B - A + \frac{C}{2}\right)' + (d A + \tilde d B)' \left(B - A + \frac{C}{2}\right) = 0.
\ee
EOM (\ref{EOM-A}) can also be expressed as
\be\label{EOM-A-1}
\left(A - \frac{\tilde d}{2 (d + \tilde d)} C\right)'' + \frac{\tilde d + 1}{r} \left(A - \frac{\tilde d}{2 (d + \tilde d)} C\right)' + (d A + \tilde d B)' \left(A - \frac{\tilde d}{2 (d + \tilde d)} C\right)' = 0.
\ee
Combining (\ref{EOM1-1-1}) and (\ref{EOM-A-1}) gives
\be\label{EOM-B-1}
\left(B + \frac{ d}{2 (d + \tilde d)} C\right)'' + \frac{\tilde d + 1}{r} \left(B + \frac{d}{2 (d + \tilde d)} C\right)' + (d A + \tilde d B)' \left(B + \frac{d}{2 (d + \tilde d)} C\right)' = 0.
\ee
We further combine (\ref{EOM-A-1}) and (\ref{EOM-B-1}) to  give
\be\label{EOM-AB-1}
\left(d A  + \tilde d B\right)'' + \frac{\tilde d + 1}{r} \left(d A  + \tilde d B \right)' +{ (d A + \tilde d B)'} ^{2} = 0.
\ee
We subtract (\ref{EOM2-1}) from $2 \times$(\ref{EOM2-2}) to give
\be\label{EOM-B-2}
\left(d A  + \tilde d B\right)'' + \frac{2 \tilde d + 3}{r} \left(d A  + \tilde d B \right)' +{ (d A + \tilde d B)'} ^{2} = 0.
\ee
The above two equations must imply 
\be
(d A + \tilde d B)' = 0,
\ee
which in turn gives, noting $A (r \to \infty) = 0, B(r \to \infty) = 0$,
\be\label{SUSY1}
d A + \tilde d B  = 0.
\ee
As we will demonstrate, this condition and the so-called ``no-force'' condition (\ref{CAP}), when combined with the other EOM's, already imply the preservation of 1/2 SUSY. The ``no-force'' condition 
(\ref{CAP}) plays the key role here. We will see later when this condition is dropped, we can find so-called Non-SUSY brane solutions.

With this, we have from (\ref{EOM2-1}) the non-trivial solution
\be
\left(A - \frac{\tilde d}{2 (d + \tilde d)} C\right)' = 0,
\ee
which gives
\be\label{SolA}
A  = \frac{\tilde d}{2 (d + \tilde d)} C,
\ee
due to $A ( r \to \infty) = C(r \to \infty) = 0$.  So with (\ref{SUSY1}), this gives also
\be\label{SolB}
B =  - \frac{d}{2 (d + \tilde d)} C.
\ee

One can check that the following
\be\label{ABC-r}
A = \frac{\tilde d}{2 (d + \tilde d)} C, \quad  B =  - \frac{d}{2 (d + \tilde d)} C
\ee
solve all equations (\ref{EOM1-1}), (\ref{EOM2-1}), (\ref{EOM2-2}) and (\ref{EOM-A-1}).  So we have also from (\ref{CAP})
\be
\bar\phi = \frac{\alpha}{2} C.
\ee
In other words, all the EOMs boil down, from (\ref{EOM4-1}), to
\be
\delta^{mn}\partial_{m} \partial_{n} e^{- C} = - 2 \kappa^{2} T_{p} \, \delta^{(D - d)} (x_{\bot})
\ee
which gives the solution
\be
e^{- C} = 1 + \frac{K_{d}}{r^{\tilde d}},
\ee
where
\be
K_{d} = \frac{2 \kappa^{2} T_{p}}{\tilde d\, \Omega_{\tilde d + 1}},  \quad \Omega_{n} = \frac{2 \pi^{(n + 1)/2}}{\Gamma ((n + 1)/2)},
\ee
with $\Omega_{n}$  the volume of unity $n$-sphere.

In summary, we find the SUSY preserving configurations in diverse dimensions with  $\tilde d = D - 2 - d \ge 1 \to D \ge 3 + d \ge 4$ due to $d \ge 1$ as
\be\label{SUSY-p-brane}
d s^{2} = e^{\frac{\tilde d}{D - 2} C} \eta_{\mu\nu} d x^{\mu} d x^{\nu} + e^{- \frac{d}{D - 2} C} \delta_{mn} d x^{m} d x^{n},
\ee
where $\mu, \nu = 0, 1 \cdots p$ and $m, n = p + 1, \dots D - 1$,
\be
A_{01\cdots p} = - \left(e^{C} - 1\right), \quad \bar\phi = \frac{\alpha}{2} C,
\ee
and
\be \label{C}
e^{- C} = 1 + \frac{K_{d}}{r^{\tilde d}},
\ee
where
\be\label{K}
K_{d} = \frac{2 \kappa^{2} T_{p}}{\tilde d\, \Omega_{\tilde d + 1}},  \quad \Omega_{n} = \frac{2 \pi^{(n + 1)/2}}{\Gamma ((n + 1)/2)},
\ee
with $\Omega_{n}$ the volume of unity $n$-sphere.

\medskip
\noindent
{\bf Certain properties of the solutions:}

\medskip

The charge density carried by a p-brane moving in spacetime is given by (\ref{p-brane-current}) as
\be
j^{M_{1} \cdots M_{p + 1}} (x) = T_{p} \int d^{p + 1} \sigma \epsilon^{i_{1} \cdots i_{p + 1}} \partial_{i_{1}} X^{M_{1}} \cdots \partial_{i_{p + 1}} X^{M_{p + 1}} \delta^{(D)} \left(x - X(\sigma)\right),
\ee
where we have taken the charge density $\mu_{p} = T_{p}$.   For a static p-brane with $\sigma^{\mu} = X^{\mu}$, we have the familiar one
\be
j^{01\cdots p} (x) = T_{p} ~\delta^{(D - p - 1)} (x_{\bot}),
\ee
and the total charge carried by the brane is
\be
Z^{1\cdots p} = \int d^{D - 1} x~ j^{01\cdots p} (x) = T_{p}~ V_{p},
\ee
which gives the charge density as $Z^{1\cdots p} /V_{p} = T_{p}$.  In general,  the current density $j^{M_{1} \cdots M_{p + 1}}$ is a (p + 1)-form.

The EOM for the (p + 1)-form potential $A_{d}$  as given in (\ref{d-formEOM}) can be expressed in terms of $j^{M_{1} \cdots M_{p + 1}}$ as
 \be\label{d-formEOM-1}
 \partial_M \left(\sqrt{- g} e^{-\alpha (d)\bar\phi} F^{M M_1 \cdots M_d} \right)
  = 2 \kappa^2 j^{M_{1} \cdots M_{d}}
 \ee
 and the Bianchi identity is simply $\partial_ {\left [M_{1}\right.} F_{M_{2} \cdots M_{d + 2 }\left. \right]}= 0$.  Note that
 \bea
 d \ast \left(e^{- \alpha \bar\phi} F_{d + 1} \right) &= & \frac{1}{(D - d - 1)!} \partial_{M} \ast \left(e^{- \alpha \bar \phi} F\right)_{M_{d + 2} \cdots M_{D}} d x^{M} \wedge d x^{M_{d + 2}}  \cdots \wedge d x^{M_{D}} \nn
 &=& \frac{\varepsilon_{M_{d + 2} \cdots M_{D} M_{1} \cdots M_{d + 1}}}{(D - d - 1)! (d + 1)!} \partial_{M} \left[ \sqrt{|g|} e^{- \alpha\bar\phi} F^{M_{1} \cdots M_{d + 1} }\right]  \nn
 &\,&\times \delta^{MM_{d + 2} \cdots M_{D}}_{N_{1} N_{2} \cdots N_{D - d}} d x^{N_{1}} \wedge d x^{N_{2}}  \cdots \wedge d x^{N_{D - d}} \nn
 &=&\frac{\varepsilon_{M_{d + 2} \cdots M_{D} M_{1} \cdots M_{d + 1}}}{(D - d - 1)! (d + 1)!} \partial_{M} \left[ \sqrt{|g|} e^{- \alpha\bar\phi} F^{M_{1} \cdots M_{d + 1} }\right] \frac{(- )^{t}}{(D - d)! d!} \nn
 &\times &   \epsilon^{M  M_{d + 2} \cdots M_{D} L_{1} \cdots L_{d}} \epsilon_{N_{1} N_{2} \cdots N_{D - d} L_{1} \cdots L_{d}} d x^{N_{1}} \wedge d x^{N_{2}}  \cdots  d x^{N_{D - d}} \nn
 &=& \frac{(-)^{D - d - 1} \delta^{M L_{1} \cdots L_{d}}_{M_{1} M_{2} \cdots M_{d + 1} }}{(D - d)! d!}   \partial_{M} \left[ \sqrt{|g|} e^{- \alpha\bar\phi} F^{M_{1} \cdots M_{d + 1} }\right]\nn
 &\,& \times \varepsilon_{N_{1} N_{2} \cdots N_{D - d} L_{1} \cdots L_{d}} d x^{N_{1}} \wedge d x^{N_{2}}  \cdots  d x^{N_{D - d}} \nn
 &=& \frac{(-)^{\tilde d + 1}} {(D - d)! d!} \partial_{M} \left[ \sqrt{|g|} e^{- \alpha\bar\phi} F^{M L_{1} \cdots L_{d} }\right]  \varepsilon_{N_{1} N_{2} \cdots N_{D - d} L_{1} \cdots L_{d}} \nn
 &\,&\times d x^{N_{1}} \wedge d x^{N_{2}}  \cdots  d x^{N_{D - d}} .
     \eea

 \bea
 d \ast \left(e^{- \alpha \bar\phi} F_{d + 1} \right)  &=& \frac{(-)^{\tilde d + 1}} {(D - d)! d!} \partial_{M} \left[ \sqrt{|g|} e^{- \alpha\bar\phi} F^{M L_{1} \cdots L_{d} }\right]  \varepsilon_{N_{1} N_{2} \cdots N_{D - d} L_{1} \cdots L_{d}} \nn
 &\,&\times d x^{N_{1}} \wedge d x^{N_{2}}  \cdots  d x^{N_{D - d}} \nn
 &=& \frac{(-)^{\tilde d + 1} 2 \kappa^{2}} {(D - d)! d!}  j^{L_{1} \cdots L_{d}} \varepsilon_{N_{1} N_{2} \cdots N_{D - d} L_{1} \cdots L_{d}}  d x^{N_{1}} \wedge d x^{N_{2}}  \cdots  d x^{N_{D - d}} \nn
 &=& (-)^{\tilde d + 1} 2 \kappa^{2} \ast J_{d},
  \eea
  where in the second equality we have used (\ref{d-formEOM-1}) and $J_{d}$ used in the last equality is a tensor defined as
      \be
      J_{d} \equiv \frac{1}{d!} \frac{j_{L_{1} \cdots L_{d}}} {\sqrt{|g|}}d x^{L_{1}} \wedge \cdots d x^{L_{d}}.
      \ee
  So writing in terms of differential forms, we have the EOM and Bianchi identity as
  \be\label{e-charge}
  d \ast \left(e^{-\alpha \bar \phi} F_{d + 1} \right) = (-)^{\tilde d + 1} 2\kappa^{2} \ast J_{d}, \qquad d F_{d + 1} = 0.
  \ee
In deriving the above, we have used the following conventions for differential forms and the Hodge duality. 
We define the totally anti-symmetric symbol $\varepsilon^{i_{1} \cdots i_{D}}$, a tensor density with weight $- 1$,  to be the same in all the frames with $\varepsilon^{1 \cdots D} = 1$, and define also
\be
\varepsilon_{i_{1} \cdots i_{D}} \equiv (-)^{t} \varepsilon^{i_{1} \cdots i_{D}}.
\ee
In the above, we denote $t$ as the number of the negative eigenvalues of the metric $g_{ij}$. We then have two tensors
\be
\epsilon^{i_{1} \cdots i_{D}} \equiv \frac{\varepsilon^{i_{1} \cdots i_{D}}}{\sqrt{|g|}}, \quad \epsilon_{i_{1} \cdots i_{D}} \equiv \sqrt{|g|} \varepsilon_{i_{1} \cdots i_{D}},
\ee
where the upper or lower indices are raised or lowered by the metric or its inverse, and $|g|$ denotes the absolute value of the metric determinant.  We define
\be
\epsilon^{i_{1} \cdots i_{D}} \epsilon_{j_{1} \cdots j_{D}} = D! (-)^{t} \delta^{i_{1} \cdots i_{D}}_{j_{1} \cdots j_{D}},
\ee
and in general ($p + q  = D$)
\be
\epsilon^{i_{1} \cdots i_{q} k_{1}\cdots k_{p}} \epsilon_{j_{1} \cdots j_{q} k_{1} \cdots k_{p}} = p! q! (-)^{t} \delta^{i_{1} \cdots i_{q}}_{j_{1} \cdots j_{q}},
\ee
\be
\delta^{i_{1} \cdots i_{q}}_{j_{1} \cdots j_{q}} \equiv \delta^{\left[i_{1} \cdots i_{q}\right]}_{\left [j_{1} \cdots j_{q}\right]}.
\ee
So we have
\be
A_{i_{1} \cdots i_{p}} \delta^{i_{1} \cdots i_{p}}_{j_{1} \cdots j_{p}} = A_{j_{1} \cdots j_{p}}.
\ee
A p-form is defined
\be
\omega_{p} = \frac{1}{p!} \omega_{i_{1} \cdots i_{p}} d x^{i_{1}} \wedge \cdots \wedge d x^{i_{p}},
\ee
with the Hodge dual basis
\be
\ast \left(d x^{i_{1}} \wedge \cdots \wedge d x^{i_{p}}\right) \equiv \frac{1}{q!} \epsilon_{j_{1} \cdots j_{q}}\,^{i_{1} \cdots i_{p}} d x^{j_{1}} \wedge \cdots \wedge d x^{j_{q}}.
\ee
We then have
\bea
\ast (\omega_{p}) &=& \frac{1}{p! q!} \epsilon_{j_{1} \cdots j_{q}}\,^{i_{1} \cdots i_{p}}  \omega_{i_{1} \cdots i_{p}} d x^{j_{1}} \wedge \cdots \wedge d x^{j_{q}}\nn
&=& \frac{1}{q!} (\ast \omega)_{j_{1} \cdots j_{q}}  d x^{j_{1}} \wedge \cdots \wedge d x^{j_{q}},
\eea
where
\be
(\ast \omega)_{j_{1} \cdots j_{q}} = \frac{1}{p!} \epsilon_{j_{1} \cdots j_{q}}\,^{i_{1} \cdots i_{p}}  \omega_{i_{1} \cdots i_{p}}.
\ee
Further
\bea
(\ast \ast \omega)_{i_{1} \cdots i_{p}} &=& \frac{1}{p! q!} \epsilon_{i_{1} \cdots i_{p}}\,^{j_{1} \cdots j_{q}} \epsilon_{j_{1} \cdots j_{q}}\,^{i'_{1} \cdots i'_{p}}  \omega_{i'_{1} \cdots i'_{p}}\nn
&=&  \frac{(-)^{pq}}{p! q!} \epsilon^{j_{1} \cdots j_{q}}\,_{i_{1} \cdots i_{p}}\epsilon_{j_{1} \cdots j_{q}} \,^{i'_{1} \cdots i'_{p}}  \omega_{i'_{1} \cdots i'_{p}}\nn
&=& (-)^{pq + t} \omega_{i_{1} \cdots i_{p}}.
\eea
Given the above, it is clear that the charge per unit p-brane volume is
\bea
e_{p} &=& \int d^{\bot} x~ j^{01\cdots p} (x_{\bot})\nn
&=& (-)^{(D - d) (d + 1)} \int \ast J_{d}\nn
&=& \frac{(-)^{D(d + 1)}}{2 \kappa^{2}} \int d \ast (e^{-\alpha\bar\phi} F_{d + 1})\nn
&=&  \frac{(-)^{D(d + 1)}}{2 \kappa^{2}} \int_{S^{\tilde d + 1} (r \to \infty)} [\ast (e^{- \alpha\bar \phi} F_{d + 1})]_{\tilde d + 1}
\eea
Let us give an evaluation of the above charge density for the SUSY solution  found earlier. Note that
\be
F_{r 01 \cdots p} = - \partial_{r} e^{C} = e^{2 C} \partial_{r} e^{- C} = - e^{2 C} \frac{K_{d} \tilde d}{r^{\tilde d + 1}},
\ee
\bea
\ast (e^{- \alpha\bar\phi} F) &=& e^{- \alpha\bar\phi} \epsilon_{\theta_{1} \cdots \theta_{D - d - 1}}\,^{r01\cdots p} F_{r01\cdots p} d\theta^{1} \wedge\cdots \wedge d\theta^{\tilde d + 1}\nn
&=& - (-)^{D (d + 1)} e^{- \alpha\bar\phi} \frac{e^{2 B (\tilde d + 1)} r^{2 (\tilde d + 1)}}{e^{A d + B (\tilde d + 2)} r^{\tilde d }} \sqrt{|g_{\Omega}|} F_{r01\cdots p} d\theta^{1} \wedge\cdots \wedge d\theta^{\tilde d + 1}\nn
&=& - (-)^{D (d + 1)} e^{-\alpha\bar\phi - A d + B \tilde d} r^{\tilde d + 1} F_{r 01\cdots p}\, d\Omega_{\tilde d + 1}\nn
&=& (-)^{D(d + 1)} e^{- \alpha\bar\phi - 2 Ad + 2 C} \tilde d K_{d}\, d\Omega_{\tilde d + 1}\nn
&=& (-)^{D(d + 1)} \tilde d K_{d}\, d\Omega_{\tilde d + 1},
\eea
where we have used $d A + \tilde d B = 0,  \alpha \bar \phi + 2 d A - 2 C = 0$.  So we have
\bea
e_{p} &=&  \frac{(-)^{D(d + 1)}}{2 \kappa^{2}} \int_{S^{\tilde d + 1} (r \to \infty)}[ \ast (e^{- \alpha\bar \phi} F_{d + 1})]_{\tilde d + 1}\nn
&=& \frac{\tilde d K_{d} \Omega_{\tilde d + 1}}{2 \kappa^{2}}\nn
&=& T_{p}
\eea
as expected.  Note that the canonical dimension for the electric-like charge $e_{p}$ is
\be
[e_{p}] = [T_{p}] = d,
\ee
while the dual magnetic charge is defined as
\be
g_{\tilde p} = \int_{S^{d + 1} (r \to \infty)} F_{d + 1},
\ee
which has canonical dimension (noting the canonical dimension for $A_{01\cdots p}$ is zero, i.e, $[A_{01\cdots p}] = 0$) as
\be
[g_{\tilde p}] = 1 - (d + 1) = - d.
\ee
In other words, the charges of these two dual objects have the opposite canonical dimension as expected such that they obey the usual Dirac charge quantization (see \cite{Nepomechie:1984wu, Teitelboim:1985ya})
\be\label{chargeq}
\frac{e_{p} g_{\tilde p}}{4 \pi} = \frac{1}{2} n,
\ee
with $n$ an integer.

However, the above definitions for the electric-like charge and the magnetic-like charge are not symmetric in the sense that the two charges are not put in the equal footing.  For example, the electric-like  charge is always given by its tension $e_{p} = T_{p}$ while the magnetic-one is given by $g_{\tilde p} \sim 2 \kappa^{2} T_{\tilde p}$.  This is not good for the electric-magnetic duality. So we redefine the respective charge as
\be
e_{p}  \equiv \frac{(-)^{D(d + 1)}}{\sqrt{2} \kappa} \int e^{- \alpha\bar\phi} [\ast(F_{d + 1})]_{\tilde d + 1} = \sqrt{2} \kappa T_{p}
\ee
while
\be
g_{\tilde p} \equiv \frac{(-)^{D(\tilde d + 1)}}{\sqrt{2}\kappa} \int F_{d + 1} = \sqrt{2} \kappa T_{\tilde p}.
\ee

Note that the good feature using the dual formulation is that we don't need to introduce the source since the $r = 0$ point is excluded from the consideration. In other words, we are using the so-called Wu-Yang construction. For this, we need to make ansatz on the field dual strength instead
\be
\tilde F_{\tilde d + 1} = e^{- \alpha\bar\phi} [\ast (F_{d +1})]_{\tilde d + 1} = (-)^{D(d + 1)} \frac{2 \kappa^{2} T_{p}}{\Omega_{\tilde d + 1}} \Omega_{\tilde d + 1},
\ee
where $\Omega_{\tilde d + 1}$ is the volume form of unit $(\tilde d + 1)$-sphere.  So the charge quantization (\ref{chargeq}) gives the tension quantization between two dual objects as
\be
2 \kappa^{2} T_{p} T_{\tilde p} = 2 \pi n.
\ee

The ADM mass per unit p-brane volume can be computed for our configuration (\ref{SUSY-p-brane}) using a special form of the general formula developed by the author \cite{Lu:1993vt} as
\bea
M_{p} &=& - \frac{\Omega_{\tilde d + 1}}{2 \kappa^{2}} \left[(\tilde d + 1) r^{\tilde d + 1} \partial_{r} e^{2 B} + (d - 1) r^{\tilde d + 1} \partial_{r} e^{2 A} \right]_{r \to \infty}\nn
&=& \frac{\Omega_{\tilde d + 1}}{2 \kappa^{2}} \tilde d K_{d} = T_{p},
\eea
again as expected. So we have the BPS bound saturated as
\be
\sqrt{2} \kappa M_{p} = e_{p}.
\ee
In general in a SUSY theory, this indicates that the underlying configuration preserves certain number of SUSY. The other indication for preserving certain underlying SUSY is via the so-called ``no-force'' condition derived earlier. Let us explore a bit further.  Consider a probe p-brane in the background found which is parallel to the source p-brane. Its dynamics along the transverse directions can be described by the following Nambu-Goto Lagrangian (for simplicity)
\bea
{\cal L}_{p} &=& - T_{p} \left[e^{d A + \alpha\bar\phi/2} \sqrt{- \det(\eta_{\alpha\beta}  + e^{2 B - 2 A} \partial_{\alpha} X^{m} \partial_{\beta} X^{m})} - (e^{C} - 1)\right]\nn
&=& - T_{p} \left[e^{d A + \alpha\bar\phi/2} \left(1 +  \frac{1}{2} e^{2 B - 2 A} \eta^{\alpha\beta}\partial_{\alpha} X^{m} \partial_{\beta} X^{m} + \cdots \right) - e^{C}  + 1\right],
\eea
where since the term $e^{d A + \alpha\bar\phi/2}$ cancels the $e^{C}$ term, the initial static probe brane will remain so if
\be
(d - 2) A + 2 B + \alpha\bar\phi/2 = C + 2 B - 2 A =  {\rm constant},
\ee
which is indeed true from (\ref{ABC-r}) with the ${\rm constant} = 0$ here. Note that the relation of $d A + \tilde d B = 0$  plays a key role in having (\ref{ABC-r}).

We now come to demonstrate explicitly using the 10D supergravities as examples that the BPS solutions found above preserve indeed one half of spacetime supersymmetries though this remains true for all solutions given above in diverse dimensions.  In addition, we will show that the zero modes associated with the p-brane configuration are the expected ones, in particular for D-branes they give the corresponding vector supermultiplet. 

\medskip
\noindent
{\bf The 10D Case}
\medskip

We first discuss the 10-dimensional case by focusing on Type IIA and Type IIB supergravities.

For the type IIA supergravity, in addition to the NSNS 2-form potential $B_{2}$, we have the so-called RR 1-form potential $A_{1}$ and the RR 3-form potential $A_{3}$. Given what we have described before, they are respectively related to the fundamental strings, the D0-branes and the D2 branes.  By the Hodge or electromagnetic duality in 10D, their respective magnetic dual objects are NSNS 5-branes, D6 branes and D4 branes.

For the type IIB supergravity, we have the same NSNS 2-form potential $B_{2}$ and for this the story remains the same as in Type IIA case. In other words, it is related to the fundamental strings and the magnetic dual objects are the NSNS 5-branes. For RR form potentials, we have here, however, the RR 0-form potential $\chi$, the RR 2-form potential $A_{2}$ and the 4-form potential $A^{+}_{4}$ whose 5-form field strength satisfies the self-duality duality relation $F_{5} = \ast F_{5}$. These RR potentials are expected to relate to the D-instantons, D1-branes(D-strings) and the self-dual D3 branes, respectively.  Their magnetic duals in 10D are D7-branes and D5-branes (Note that the D3 branes are self-dual).

The above indicate that the Dp branes in type IIA are those with even p while the Dp branes in Type IIB are those with odd p.

In the following, we will limit ourselves to those p-brane configurations  with well-defined asymptotically behavior.  In other words, we limit\footnote{The D-instanton solution (corresponding to $p = -1$) is referred to \cite{Gibbons:1995vg}.} to $0 \le p \le 6$.

In the 10D case, we will use the 1/2 BPS F-string configuration and the (anti) self-dual D3 brane configuration given above to demonstrate explicitly the preservation of 1/2 spacetime SUSY and the other cases can be done in a similar fashion, then we move to discuss the new brane-scan given earlier, finally we will discuss the zero modes for each of the brane solitons in 10D and discuss in detail for the F-string and the (anti) self-dual D3 in IIB theory as illustrations. 

\medskip
\noindent
{\sl 1/2 SUSY preservation: F-string as an example}
\medskip

 This 1/2 BPS F-string solution was first given in \cite{Dabholkar:1990yf}.  For showing the configuration (\ref{SUSY-p-brane}) preserving one half of the spacetime SUSY, we don't need the explicit solution. All we need are the following relations
 \be\label{ABC-phi}
 A = \frac{\tilde d}{2 (d + \tilde d)} C, \quad B = - \frac{d}{2 (d + \tilde d)} C,  \quad \bar \phi = \frac{\alpha (d)}{2} C,
 \ee
where $A, B$  are two functions given in the metric (\ref{pmetric}) which for convenience is given below
 \be\label{p-brane-metric}
ds^{2} = e^{2 A(r)} \eta_{\mu\nu} d x^{\mu} d x^{\nu} +  e^{2 B(r)} \delta_{mn} d x^{m} d x^{n},
\ee
and $C$ determines the form potential as given in (\ref{d-formp}).

We now specify to the F-string configuration for which we have $d = 2, \tilde d = 6$ in 10D.   As explained earlier, for this static vacuum-like configuration, we need to set all the fermion fields vanish. Given that under SUSY transformations, the variations of bosonic fields are directly related to the fermionic ones and so we have automatically $\delta_{\rm SUSY}\,g_{MN} = 0, \delta_{\rm SUSY}\,B_{MN} = 0, \delta_{\rm SUSY}\, \phi = 0$. In other words, the F-string configuration, which has $P_{2} \times SO(8)$ symmetry and involves only bosonic fields with respect to this symmetry, is invariant under the underlying SUSY.  To actually have certain unbroken SUSY, we need to show that there exists certain number of Killing spinors under which the fermionic fields will remain to vanish under the corresponding SUSY transformations. This is not obvious at a first look since we have non-vanishing bosonic fields for this configuration.  As we will show below, this F-string configuration preserves one half of the spacetime SUSY.

In other words,  we need to show that there exist 16 Killing spinors $\epsilon$ under which the transformations of the gravitino $\delta_{\rm SUSY}\, \psi_{M}$ and the dilatino $\delta_{\rm SUSY}\, \lambda$ also vanish, i.e.
 \bea\label{Killing-spinor-eq}
 \delta_{\rm SUSY}\, \psi_{M} &=& D_{M}\, \epsilon - \frac{1}{96} e^{- \bar \phi/2} \left(\Gamma_{M}\,^{NPQ} - 9 \delta_{M}\,^{N} \Gamma^{PQ}\right)H_{NPQ} \,\Gamma^{11} \,\epsilon = 0,\nn
 \delta_{\rm SUSY}\, \lambda & =& \frac{1}{3}  \Gamma^{M} \partial_{M}\bar \phi\,\Gamma^{11} \epsilon - \frac{1}{36} e^{-\bar \phi/2} \Gamma^{MNP} H_{MNP}\, \epsilon = 0,
 \eea
 where $H_{MNP} = 3 \partial_{\left[M\right.} B_{\left. NP\right]}$ and the covariant derivative is
 \be
 D_M \epsilon = \left(\partial_M - \frac{1}{4} \omega_{M AB} \Gamma^{AB}\right) \epsilon,
 \ee
 with the spin-connection defined as
 \be
 \omega_{MNP} = \Omega_{MNP} + \Omega_{PNM} + \Omega_{PMN},
 \ee
with $\Omega_{MNP} = e^A_P \partial_{\left[M\right.} e_{\left. N\right] A}$ in the case of vanishing gravitino.  In the above, $M, N, P, \cdots $ stand for the spacetime curved indices while $A, B, \cdots$ stand for the flat Lorentz indices. $e^A_M$ is the zehnbein.  

With the metric (\ref{p-brane-metric}),  note that
   \be
   \omega_{MAB} \Gamma^{AB} = - \Omega_{NPM} \Gamma^{NP} + 2~ \Omega_{MNP} \Gamma^{NP},
   \ee
   we have
   \bea\label{spin-con}
   \omega_{\mu AB} \Gamma^{AB} &=& - 2 \,\partial_n A \,\Gamma^n\,_\mu,\nn
   \omega_{m AB} \Gamma^{AB} &=& -  2\, \partial_n B\, \Gamma^n\,_m.
   \eea
  $\Gamma^A$ are the 10D Dirac matrices satisfying
  \be\label{direcg}
  \{\Gamma^A, \Gamma^B\} = 2 \eta^{AB},
  \ee
  with $\eta^{AB} = (- 1, 1, \cdots 1)$.    $\Gamma^{AB \cdots C} \equiv \Gamma^{\left[A\right.} \Gamma^B \cdots \Gamma^{\left. C\right]}$, for example,
  \be
  \Gamma^{AB} = \frac{1}{2} \left(\Gamma^A \Gamma^B - \Gamma^B \Gamma^A\right).
  \ee
  Note that $\Gamma^{11} \equiv \Gamma^{0} \Gamma^{1} \cdots \Gamma^{9}$ with $(\Gamma^{11})^{2} = \mathbb{I}_{32\times 32}$ with $ \mathbb{I}_{N\times N}$ the $N \times N$ unit matrix.  The $\Gamma$'s with spacetime indices $M, N, P, \cdots$ have been converted using zehnbein $e_M^A$.

For the 1/2 BPS F-string, we make a 2/8 split
  \be
  \Gamma^A = \left(\rho^\alpha \otimes \mathbb{I}_{16\times 16},\,\rho^2
\otimes \gamma^a  \right),
\ee
where $\rho^\alpha$ and $\gamma^a$ are the $D = 2$  and the Euclidean $D = 8$ Dirac matrices, respectively. Here we  define
\be
\rho^2 = \rho^0 \rho^1,   \to (\rho^2)^2 = \mathbb{I}_{2\times 2}.
\ee
and
\be
\gamma^9 = \gamma^1 \gamma^2 \cdots \gamma^8, \to (\gamma^9)^2 = \mathbb{I}_{16\times 16}.
\ee
So
\be
\Gamma^{11} = \Gamma^{0} \Gamma^{1} \cdots \Gamma^{9} = \rho^{2} \otimes \gamma^{9}.
\ee
 The most general spinor, consistent with the $P_2 \times SO(8)$ symmetry, takes the form\footnote{To simplify notations, we use $x$ to represent $x^{\mu}$, along the brane directions, while $y$ to represent 
 $x^{m}$, orthogonal to the brane.}
  \be
  \epsilon (x, y) = \varepsilon (x) \otimes \eta (y),
  \ee
  where $\varepsilon (x)$ is a SO(1, 1) spinor while $\eta (y)$ is a SO(8) spinor.   Noting that the only non-vanishing components of $H_{MNP}$ are
  \be
  H_{m01} = - \partial_m e^C,
  \ee
  then the Killing spinor equations (\ref{Killing-spinor-eq}) for the present case become
  \bea\label{Killing-spinor}
  &&\Gamma^n \rho^{2} \varepsilon (x) \otimes  \left(\partial_n \bar\phi \gamma^{9} + \frac{1}{2} e^{- \frac{\bar\phi}{2} - 2 A + C} \partial_n C \right) \eta (y) = 0 \, (\leftarrow \delta \lambda = 0), \nn
  && \partial_{\mu} \varepsilon (x) \otimes \eta (y) + \frac{1}{2} \Gamma^{n}\,_{\mu} \varepsilon(x) \otimes \left(\partial_{n} A + \frac{3}{8} e^{- \bar\phi/2 - 2 A + C} \partial_{n} C ~\gamma^{9}\right)\eta (y) = 0 \,(\leftarrow \delta \psi_{\mu} = 0),\nn
  && \varepsilon(x) \otimes \left(\partial_{m} \eta (y) - \frac{3}{16} e^{- \bar\phi/2 - 2 A + C} \partial_{m} C \gamma^{9} \eta (y) \right) \nn
  && + \frac{1}{2} \Gamma^{n}\,_{m} \left(\partial_{n} B - \frac{1}{8} e^{- \bar \phi/2 - 2A + C} \partial_{n} C \gamma^{9}\right)\epsilon (x, y)  = 0\, (\leftarrow \delta \psi_{m} = 0).
  \eea
  Note that for this F-string configuration, we have $d = 2, \tilde d = 6$, $\alpha(2) = 1$ and from (\ref{ABC-phi})  the following
  \be
    A = \frac{3}{8}~ C, \quad B = - \frac{1}{8}~ C, \quad \bar\phi = \frac{1}{2} ~C.
  \ee
 With these, the first equation in (\ref{Killing-spinor}) reduces to $(1 + \gamma^{9}) \eta (y) = 0$,  the second equation gives $\varepsilon (x) = \varepsilon_{0}$, i.e. a constant 2D spinor, and the third equation gives
  $ \eta (y) = e^{- 3 C/16} \eta_{0}$ with $\eta_{0}$ a constant 8D spinor satisfying $(1 + \gamma^{9}) \eta_{0} = 0$. 
  The above says that the F-string configuration has the following
  \be
  \epsilon (x, y) = e^{- 3 C/16} \varepsilon_{0} \otimes \eta_{0},
  \ee
  with $\eta_{0}$ satisfying
  \be\label{kspinor}
   \qquad (1 + \gamma^{9})\eta_{0} = 0.
  \ee
  For Type IIA,  the SUSY spinor parameter $\epsilon (x, y)$ is a Majorana one, therefore having 32 real components. If we decompose it with respect to $SO(1, 1) \times SO(8)$ as $\epsilon (x, y) = \varepsilon (x) \otimes \eta (y)$ with $\varepsilon(x)$ and $\eta(y)$ the respective Majorana spinors in 2 dimensions and 8 dimensions. The former has 2 real components while the latter has 16 real components, giving a total of 32 real components in general. For this F-string configuration, (\ref{kspinor}) implies $(1 + \gamma^{9}) \eta (y) = 0$.  This in turn implies that only half of the 16 real components of the spinor $\eta (y)$ satisfying $(1 + \gamma^{9})\eta = 0$ will leave the F-string configuration invariant under the corresponding SUSY transformations (The other half will not) since $(1 + \gamma^{9})/2$ is a projection operator (Noting that ${\rm Tr} \,\gamma^{9} = 0, (\gamma^{9})^{2} = \mathbb{I}_{16\times 16}$, so $\gamma^{9}$ has eight ``$+ 1$'' eigenvalues and eight ``$- 1$'' eigenvalues). In other words, we have a total of $2 \times 8 = 16$ real components of $\epsilon (x, y) = e^{- 3 C(r)/16} \varepsilon_{0} \otimes \eta_{0}$ or 16 Killing spinors, which leave the F-string configuration invariant.  So this F-string configuration is a 1/2 BPS one.
  
The broken half SUSY become Goldstinos which are the 16 off-shell fermionic zero modes of this 1/2 BPS F-string, giving 8 on-shell fermionic ones in addition to the 8 on-shell bosonic ones.
Strictly speaking, we need also to show that the 1/2 BPS F-string being initially static will remain so and this can be shown using its sigma-model action by setting vanish all the fermionic coordinate $\theta = 0$. This action is, in Einstein frame,
\be
S_{1} =   - \frac{T_{1}}{2} \int d^{2} \sigma \left(\sqrt{ - h} h^{\alpha\beta} \partial_{\alpha} X^{M} \partial_{\beta} X^{N} e^{\bar \phi/2} g_{MN} (X) +  \epsilon^{\alpha\beta} \partial_{\alpha} X^{M} \partial_{\beta} X^{N} B_{MN} (X) \right),
\ee
where $h_{\alpha\beta}$ is the induced metric given by $h_{\alpha\beta} = \partial_{\alpha} X^{M} \partial_{\beta} X^{N} e^{\bar\phi/2} g_{MN}$, up to a scaling function $f (\sigma)$ in this case, with $g_{MN}$ and $B_{MN}$ along with the dilaton $\bar\phi$ given by the corresponding solution. Here $\alpha, \beta = 0, 1$.
The string EOM is
\bea\label{string-EOM}
&&\partial_{\alpha}\left(\sqrt{- h} h^{\alpha\beta} \partial_{\beta} X^{N} g_{MN} e^{\bar\phi/2}\right) - \frac{1}{2} \sqrt{- h} h^{\alpha\beta} \partial_{\alpha} X^{N} \partial_{\beta} X^{P} \partial_{M} \left(g_{NP} e^{\bar\phi/2}\right)\nn
&&- \frac{1}{2} \epsilon^{\alpha\beta} \partial_{\alpha} X^{N} \partial_{\beta} X^{P} H_{MNP} = 0.
\eea
For this infinity long string, we choose the static gauge $X^{\mu} = \sigma^{\mu}$ with $\mu = \alpha = 0, 1$. Note that
\be
h_{00} = - h_{11} = - e^{2 A + \bar\phi/2} = -  e^{C}, \quad \sqrt{- h} = e^{2A + \bar\phi/2} = e^{C},
\ee
When $M = \mu$, the EOM satisfies trivially and when $M = m$, we have
\be
\partial_{0}^{2} X^{m} = 0,
\ee
which just says that the F-string remains static if initially being so.  This is called ``no-force'' condition mentioned earlier. 

 Again for the worldsheet field $X^{M}$ and $\theta$, we need to set $\theta = 0$ for the 1/2 BPS F-string,  just like the bulk case, for which under SUSY $\delta_{\rm SUSY}\, X^{M} = 0$ but for $\theta$ we need also $\delta_{\rm SUSY}\, \theta = 0$. 
 In general, we have $\delta_{\rm SUSY}\, \theta = \epsilon$ with $\epsilon$ a 10D Majorana spinor. But  we also have the so-called $\kappa$ symmetry which can be used to gauge away half of the $\theta$ such that we have only half of the $\epsilon$ left which has to be set vanish for this 1/2 BPS F-string configuration. In other words, this string is again preserving one half of the spacetime SUSY which are just those gauged away by the $\kappa$ symmetry (total 16). While those broken are just those Goldstinos, i.e., the fermionic zero modes.  So everything is consistent.  
  
 Due to the ``no-force'' condition, we  have multiple string solutions placed at different locations so long they are parallel to each other. In other words, given a single center solution (\ref{SUSY-p-brane})-(\ref{K}) for a 1/2 BPS F-string, the following multi-center one also solves all the EOM and preserves one half of spacetime SUSY,
 \be\label{multi-center-sol}
 e^{- C} =  1 + \sum_{l} \frac{K^{(l)}_{2}}{|\vec{r}  - \vec{r}_{l}|^{6}}.
 \ee
 Finally, we come to count the zero modes for the 1/2 BPS F-string. 

\medskip
\noindent
{\sl Zero-modes:}
\medskip

Due to the solution, we spontaneously break the translational symmetries along the directions transverse to the F-string, therefore giving rise to 8 translational zero modes $x^{m}$ ($m = 2, \cdots 9$).  In addition, this solution breaks one half of spacetime supersymmetries, i.e., 16 off shell-modes, counting 8 on-shell fermionic zero modes, giving a total $8_{B} + 8_{F}$, as expected. This is the case for both IIA and IIB.

Actually the same F-string solution solves also the respective EOM for  the heterotic cases for which we have only $N = 1$ 10D SUSY (Note that for the Type I case, we don't have  a stable F-string but we have a 1/2 BPS D-string solution). For the heterotic cases, we do have 1/2 BPS F-strings. 
   
For the heterotic cases we have only one supersymmetric mover, either left or right, we therefore only count half of the bosonic translational zero modes $x^{m}$, giving $8/2 = 4$ zero-modes.  We have also 4 fermionic on-shell zero modes, giving a total of $4_{B} + 4_{F}$, the expected result.
 
  In the above, we give a complete discussion of 1/2 BPS F-string in IIA (This also gives 1/2 BPS F-string in Type IIB).  By the same token, we can also show that the other p-brane solutions given in (\ref{SUSY-p-brane}) each preserves one half of the spacetime SUSY using the respective Killing spinor equations, which can be obtained from the general IIA SUSY transformations for both gravitino $\psi_{M}$ and the dilatino $\lambda$ given below by focusing on a given form field strength once a time (set all other form field strengths to vanish) and set also both $\delta\psi_{M}$ and $\delta_{\rm} \lambda$ to vanish.  These general SUSY transformations for IIA are given here as
\bea\label{IIA-K-Spinore-E1}
\delta \psi_{M} &=& D_{M} \epsilon + \frac{1}{64} e^{3\bar\phi/4} \left(\Gamma_{M}\,^{NP} - 14 \delta_{M}^{N} \Gamma^{P}\right) F_{NP} \Gamma^{11} \, \epsilon\nn
&\,& - \frac{1}{96} e^{-\bar\phi/2} \left(\Gamma_{M}\,^{NPQ} - 9 \delta_{M}^{N} \Gamma^{PQ}\right) H_{NPQ}\Gamma^{11}\, \epsilon\nn
&\,& - \frac{1}{256} e^{\bar\phi/4} \left(\Gamma_{M}\,^{NPQR} - \frac{20}{3} \delta_{M}^{N} \Gamma^{PQR}\right) \,\tilde F_{NPQR} \epsilon, \nn
\delta \lambda &=& \frac{1}{3} \Gamma^{M} \partial_{M} \bar\phi \Gamma^{11} \epsilon + \frac{1}{8} e^{3 \bar\phi/4} \Gamma^{MN} F_{MN} \epsilon\nn
&\,& - \frac{1}{(3!)^{2}} e^{- \bar\phi/2} \Gamma^{MNP} H_{MNP}\, \epsilon - \frac{1}{12 \cdot 4!} e^{\bar\phi/4} \Gamma^{MNPQ} \tilde F_{MNPQ} \Gamma^{11} \epsilon,
\eea
 where some notations are given in the discussion of 1/2 BPS F-string and here
 \be
 \tilde F_{4} = d A_{3} + A_{1} \wedge H_{3}.
 \ee

One can use the respective Killing spinor equations from (\ref{IIA-K-Spinore-E1})  to show that the $p = 0, 2, 4, 6$  solutions indeed preserve one half of the corresponding spacetime SUSY. For this, we need to consider each given form field strength at one time in  (\ref{IIA-K-Spinore-E1}). For examples, for $p = 0$, we need to consider only the electric-like 2-form $F_{m 0} = - \partial_{m} e^{C}$ in the above but for $p = 6$, we need to replace the two form $F_{2}$ using its magnetic-dual, i.e. $F_{2} = e^{- \alpha (7)\bar\phi} \ast F_{8}$,  with $F_{m012\cdots 6} = - \partial_{m} e^{C}$ to check the one half of SUSY preservation. Similarly for D2 and its magnetic dual D4 as well as for the magnetic dual of F-string, i.e., the NSNS 5-brane case.

For IIB case, we need to use the following Killing spinor equations from the respective $\delta \psi_{M}$ and $\delta \lambda$ given in \cite{Schwarz:1983qr}.  Here we adopt a better form which is much more convenient as

\bea\label{IIB-Killing-Spinor-E}
\delta \lambda &=& \frac{1}{2}\left(\partial\bar\phi - i e^{\bar\phi} \partial_{M} \chi\right) \Gamma^{M} \epsilon  +  \frac{1}{4} e^{- \bar\phi/4} \left(i e^{\bar\phi} \tilde F^{(3)} - H^{(3)} \right) \epsilon^{*}, \nn
\delta \psi_{M} &=& D_{M} \epsilon - \frac{i}{4} e^{\bar\phi} \partial_{M} \chi \epsilon - \frac{i}{16} \tilde F^{(5)} \Gamma_{M} \epsilon \nn
                        &\,& - \frac{1}{96} e^{- \bar\phi/2} \left(\Gamma_{M}\,^{NPQ} - 9 \delta_{M}^{N} \Gamma^{PQ} \right) H_{NPQ} \epsilon^{*} \nn
                        &\,& - \frac{i}{96} e^{\bar\phi/2} \left(\Gamma_{M}\,^{NPQ} - 9 \delta_{M}^{N} \Gamma^{PQ} \right) \tilde F_{NPQ} \epsilon^{*},
 \eea
where both $\psi_{M}$ and $\lambda$ are Majorana-Weyl spinors satisfying $\Gamma^{11} \psi_{M} = \psi_{M}$ and $\Gamma^{11} \lambda = - \lambda$.   So is the spinor $\epsilon$ satisfying $\Gamma^{11} \epsilon = \epsilon$. In the above, we denote $\epsilon^{*}$ as the complex conjugate of $\epsilon$. $\Gamma^{11}$ is the product of the ten Dirac $\Gamma^{A}$. Other notations are the same as given in the above for IIA.  We also have the following
\be
F^{(n)} \equiv \frac{1}{n!} \Gamma^{M_{1} \cdots M_{n}} F_{M_{1} \cdots M_{n}},
\ee
and
\bea
F_{3} &=& d A_{2}, \quad H_{3} = d B_{2}, \quad F_{5} = d A_{4},\nn
\tilde F_{3} &=& F_{3} - \chi H_{3}, \nn
\tilde F_{5} &=& F_{5} - \frac{1}{2} \left(A_{2} \wedge H_{3} - B_{2} \wedge F_{3}\right).
\eea
  In addition, the 5-form $\tilde F_{5}$ satisfies the following anti self-duality relation\footnote{In the original Schwarz's paper \cite{Schwarz:1983qr}, we have the self-duality relation $\tilde F_{5} = \ast \tilde F_{5}$ for which the signature is $(+, -, \cdots, -)$. When we change the signature to $(-, +, \cdots, +)$, the self-duality becomes an  anti self-dual one which can be seen easily as follows. For simplicity, we take the spacetime flat and the self-duality in the original signature is
  \be 
 \tilde F_{A_{1} \cdots A_{5}} = \frac{1}{5!} \epsilon_{A_{1} \cdots A_{5}}\,^{B_{1} \cdots B_{5}} \tilde F_{B_{1} \cdots B_{5}},
  \ee
  and note $\epsilon^{01\cdots 9} = 1$ and $\epsilon_{A_{1} \cdots A_{5}}\,^{B_{1} \cdots B_{5}} = \eta_{A_{1} A'_{1}} \eta_{A_{2}A'_{2}} \cdots \eta_{A_{5}A'_{5}} \epsilon^{A'_{1} A'_{2} \cdots A'_{5} B_{1} \cdots B_{5}}$ with 5 $\eta_{AB}$'s.  So when we change the signature which amounts to sending  $\eta_{AB} \to - \eta_{AB}$, we have $\epsilon_{A_{1} \cdots A_{5}}\,^{B_{1} \cdots B_{5}} \to - \epsilon_{A_{1} \cdots A_{5}}\,^{B_{1} \cdots B_{5}} $, therefore the self-duality becomes an anti self dual one. 
  
    }
  \be
\tilde F_{5} = - \ast \tilde F_{5}.
\ee
In the above, $\chi$ is zero form axion in IIB and the (anti) self-dual 5-form $\tilde F_{5}$ is related to the dyonic D3 in this theory which we will come to give its explicit example in what follows.
   
\medskip
\noindent
{\sl The 1/2 BPS D3 brane:}
\medskip

This (anti) self-dual 1/2 BPS soliton solution was given a while ago by Duff and myself \cite{Duff:1991pea}. For this case, from our general solutions (\ref{SUSY-p-brane}), we have, noting $d = \tilde d = 4$,
\be\label{d3}
A = - B = \frac{C}{4}, \quad \bar\phi = 0,
\ee
where the last equality comes from $\alpha (4) = 0$ which implies that the dilaton is a constant. In the present context, we  have only form field strength $F_{5} \neq 0$. The Killing spinor equations for the present case is, from (\ref{IIB-Killing-Spinor-E}),
\bea\label{D3-Killing-Spinor-E}
\delta_{\rm SUSY}\, \psi_{M} &=& D_{M} \epsilon - \frac{i}{4 \times 480} \Gamma^{M_{1} M_{2} \cdots M_{5}} F_{M_{1} M_{2} \cdots M_{5}} \Gamma_{M} \epsilon = 0,\nn
\delta_{\rm SUSY}\, \lambda &=&  0.
 \eea
The above Killing spinor equation $\delta_{\rm SUSY}\, \lambda = 0$ satisfies automatically due to $\bar\phi = 0$ and all other form field strengths being zero in its expression. To check the Killing spinor equations for $\delta _{\rm SUSY}\, \psi_{M} = 0$, we need first to solve the anti self-dual relation for $\tilde F_{5}$. Given $F_{m0123} = - \partial_{m} e^{C}$, we have
\be
F_{m_{1}\cdots m_{5}} = -\frac{\varepsilon_{m_{1} \cdots m_{5}}\,^{m 0123}}{\sqrt{- g}} F_{m0123} =  \varepsilon^{m_{1} \cdots m_{5} m} e^{- C} \partial_{m} C ,
\ee
where we have used $\sqrt{- g} = e^{6B + 4 A}$ and the relations given (\ref{d3}).  Here $\varepsilon^{m_{1} \cdots m_{6}}$ is total antisymmetric with respective to the transverse indices and with $\varepsilon^{456789} = 1$.

For the present case, we need to make a 4/6 split on the indices $M = (\mu, m)$ and $A = (\alpha, 3 + a)$ with $\mu = 0, 1, 2, 3$, $m = 4, 5, \cdots 9$; and $\alpha = 0, 1, 2, 3$, $a = 1, 2, \cdots 6$. Then we have the Dirac matrices $\Gamma^{A}$
\be
\Gamma^{A} = \left(\gamma^{\alpha} \otimes \mathbb{I}_{8 \times 8}, \Gamma^{3 + a} = \gamma^{5} \otimes \Sigma^{a}\right),
\ee
where $\gamma^{\alpha}$ are the usual 4D Dirac matrices, $\gamma^{5} = - i \gamma^{0} \gamma^{1} \gamma^{2} \gamma^{3}$ with $(\gamma^{5})^{2} = \mathbb{I}_{4 \times 4}$, and $\Sigma^{a}$ are the Euclidean 6D Dirac matrices with the following
\be
\Sigma^{7} = i \, \Sigma^{1} \cdots \Sigma^{6}, \quad (\Sigma^{7})^{2} = \mathbb{I}_{8 \times 8}.
\ee
So we have
\be
\frac{1}{5!} \Sigma^{a_{1} \cdots a_{5}} \varepsilon_{a_{1} \cdots a_{5} a} = - i \Sigma^{7} \Sigma_{a}, 
\ee
which gives
\bea
\Gamma^{M_{1} \cdots M_{5}} F_{M_{1} \cdots M_{5}} &=& 5! \Gamma^{m 0123} F_{m0123} +  \Gamma^{m_{1} \cdots m_{5}} F_{m_{1} \cdots m_{5}}\nn
&=& - i\, 5! \left(\mathbb{I} + \Gamma^{11}\right) \gamma^{5} \otimes \mathbb {I}_{8 \times 8}  \Gamma^{m} \partial_{m} C,
\eea
where we have used $4A - C = 4 B + C = 0$. So we have
\bea 
\Gamma^{M_{1} \cdots M_{5}} F_{M_{1} \cdots M_{5}} \Gamma_{M} \epsilon &=& - 5! \, i \left(\mathbb{I}_{32\times 32} + \Gamma^{11} \right) \gamma^{5} \otimes \mathbb{I}_{8\times 8} \Gamma^{m} \partial_{m} C \Gamma_{M} \epsilon, \nn
&=& - 2\cdot \,5! \, i \gamma^{5} \otimes \mathbb{I}_{8\times 8} \Gamma^{m} \partial_{m} C \Gamma_{M} \epsilon,
\eea
where we have used $\Gamma^{11} \epsilon = \epsilon$ in the last equality as given below (\ref{IIB-Killing-Spinor-E}). From the first Killing equation in (\ref{D3-Killing-Spinor-E}) with the computations given in (\ref{spin-con}), we have

\bea
\partial_{\mu} \epsilon + \frac{1}{8} \partial_{n} C\, \Gamma^{n}\,_{\mu} - \frac{1}{8} \gamma^{5} \otimes \mathbb{I}_{8 \times 8} \,\Gamma^{n} \partial_{n} C\, \Gamma_{\mu} \,\epsilon &=& 0\quad (\leftarrow \delta \psi_{\mu} = 0), \nn
 \partial_{m} \epsilon - \frac{1}{8} \partial_{n} C \,\Gamma^{n}\,_{m}  - \frac{1}{8}  \gamma^{5} \otimes \mathbb{I}_{8 \times 8}\, \Gamma^{n} \partial_{n} C\, \Gamma_{m}\, \epsilon &=& 0\quad (\leftarrow \delta \psi_{m} = 0), \nn
\eea
where we have used (\ref{spin-con}) and $A = C/4,  B = - C/4$. Note that $\Gamma^{n} \Gamma_{\mu} = \Gamma^{n}\,_{\mu}$ and $\Gamma^{n} \Gamma_{m} = \delta^{n}_{m} + \Gamma^{n}\,_{m}$. So the above can  further be written as
\bea
\partial_{\mu} \epsilon + \frac{1}{8} \Gamma^{n}\,_{\mu}\partial_{n} C \left(1  + \gamma^{5}\right) \epsilon &=& 0, \nn
 \partial_{m} \epsilon -\frac{1}{8} \gamma^{5}\partial_{m} C\, \epsilon  -  \frac{1}{8} \partial_{n} C \Gamma^{n}\,_{m} \left(1  +  \gamma^{5} \right) \epsilon &=& 0.
\eea
If we express the SUSY parameter spinor $\epsilon (x^{\mu}, x^{m}) = \varepsilon (x^{\mu}) \otimes \eta (x^{m})$ with $\varepsilon (x^{\mu})$ the SO(1, 3)spinor and $\eta (x^{m})$ the SO(6) spinor, from the above we have $\varepsilon (x^{\mu}) = \varepsilon_{0}$ a constant spinor and $\eta(x^{m}) = e^{ - C/8} \eta_{0}$ with $\eta_{0}$ also a constant spinor, both of which satisfy
\be
\left(1 + \gamma^{5}\right) \varepsilon_{0} = 0,  \quad \left(1 + \Sigma^{7}\right) \eta_{0} = 0,
\ee 
where the second one is correlated with the first one due to $\Gamma^{11} \epsilon = \gamma^{5} \varepsilon \otimes \Sigma^{7} \eta = \epsilon$.
So we have
\be
\epsilon (x, y) = e^{- C (r)/8} \varepsilon_{0} \otimes \eta_{0}, \quad \left(1 + \gamma^{5} \right) \varepsilon_{0} = 0, \quad \left(1 + \Sigma^{7}\right) \eta_{0} = 0.
\ee
So by the same token as in F-string in IIA case analyzed earlier, this (anti) self-dual D3 brane configuration preserves also one half of the spacetime SUSY.

This (anti) self-dual 1/2 BPS D3 carries equal electric-like and magnetic-like charges, as indicated by the (anti) self-duality of the 5-form field strength, therefore it is a dyonic object. We now come to count its zero modes.

\medskip
\noindent
{\sl Zero-modes:}
\medskip

The 1/2 broken SUSY gives rise to 8 on-shell fermionic zero modes $8_{F}$. Here for the D3, it has 6 broken translational symmetries, giving 6 translational zero modes $x^{m}$. So unlike the F-string case, we are short of 2 bosonic zero modes due to the underlying SUSY. These two extra zero modes come from the excitation of the complex three form $G_{MNP}$ \cite{Duff:1991pea} as described by the following equation
\be
D^{P} G_{MNP} = - \frac{1}{3!} F_{MNPQR} G^{PQR},
\ee
which is solved by $b_{2} = e^{i k \cdot x} \, E\wedge d e^{2A}$ and $G_{3} = d b_{2} + i \,\ast d b_{2}$.  Here $k$ is a null vector in two Lorentzian directions tangent to the D3 worldvolume. $E$ is a constant polarization vector orthogonal to $k$ but tangent also to the worldvolume and here $\ast$ denotes the Hodge dual in the worldvolume directions (so $\ast d b_{2}$ is still a 3-form). 

 Although $G$ is complex, the zero-mode solution gives only one real vector field on the worldvolume which provides the other two zero modes needed. Together with the other zero modes, these fields make up the $d = 4, N = 4$ Super Yang-Mills supermultiplet $(A_{\mu}, \lambda^{I}, \phi^{[IJ]})$ with $I = 1, 2, 3, 4$.  This is one of nowadays so-called  1/2 BPS Dp-branes, different from the previously known ones, whose worldvolume fields consist of a vector supermultiplet rather than the scalar supermultiplet  in the old brane-scan. This turns out to be very important, lending support to the Polchinski's open string description of D-branes found via the open string T-duality as discussed at the outset, if one makes the recognition that the zero modes of the vector supermultiplet are just the massless ones of an open string in the present context. 
 
 Again by the same token, we can also show in the IIB case that each of the $p = 1, 5$ branes  preserves one half of spacetime SUSY using the SUSY transformations (\ref{IIB-Killing-Spinor-E}). For this,  we again need to consider each given form field strength at one time in  (\ref{IIB-Killing-Spinor-E}). For examples, for $p = 1$, we have the 1/2 BPS F-string if we take $H_{m01} = - \partial_{m} e^{C}$ and 1/2 BPS D-string if we take $F_{m01} = - \partial_{m} e^{C}$ while setting all other form fields to vanish. For $p = 5$, we have the 1/2 BPS NSNS 5-brane if we take $H_{3} = e^{\bar \phi} \ast H_{7}$ with $H_{m 01\cdots 5} = - \partial_{m} e^{C}$ and 1/2 BPS D5 if we take $F_{3} =  e^{- \bar \phi} \ast F_{7}$ with $F_{m01\cdots 5} = - \partial_{m} e^{C}$ while setting also the other irrelevant form field strengths to vanish. 

With the above explicit demonstrations, we hope to convince that the p-brane solutions found are indeed 1/2 BPS p-branes, each preserving one half of spacetime SUSY. We now come to give a classification of branes on the new brane scan which can be done merely based on the SUSY requirement for extended objects (see \cite{Duff:1992hu, Duff:1994an}). 

\medskip
\noindent
{\bf The brane-scan}
\medskip

For a supersymmetric p-brane moving in spacetime, it can be described by its embedding coordinates $X^{M} (\sigma)$ with $\sigma^{\alpha}$ standing for its worldvolume coordinates. Here spacetime world indices $M = 0, 1, \cdots D - 1$ and the worldvolume ones $\alpha = 0, 1, \cdots p$ with $p \le D - 1$.  Denote $\sigma^{\alpha} = (\tau, \sigma^{a})$ with $a = 1, \cdots p$. The worldvolume dimension is $d = 1 + p$.
 We can always take a ``static gauge choice" to give  $D = d + (D - d)$ split
 \be
 X^M (\sigma) = \left(X^\mu (\sigma),  Y^m (\sigma)\right)
 \ee
 with
 \be
 X^\mu (\sigma) = \sigma^\mu.
 \ee
In the above, $\mu = 0, 1, \cdots, d - 1$ and $m = d, d + 1, \cdots, D - 1$.

The physical (on-shell) worldvolume degrees of freedom (DOF) are given by $(D - d)  \, Y^m (\sigma)$ scalars.

If $Y^m (\sigma)$ are the only bosonic DOF (i.e., only scalars), we have
 \be
 N_B = D - d.
 \ee
 
A super p-brane requires in addition anti-commuting fermionic coordinates $\theta (\sigma)$. Depending on $D$, $\theta (\sigma)$ can be Dirac, Weyl, Majorana or Majorana-Weyl spinor.  As mentioned earlier, in the GS-like formalism, the fermionic $\kappa$-symmetry is a must and this eliminates half of the spinor independent components  by a physical gauge choice. 
The net result is: the theory exhibits a d-dimensional worldvolume supersymmetry with the number of fermionic generators being exactly half of the generators in the original spacetime supersymmetry.

Given this, we have the physical (on-shell) fermionic DOF
\be
N_F = \frac{1}{2} m n = \frac{1}{4} M N,
\ee
where $m$ is the independent components of a minimal spinor in the worldvolume d-dimensions and $n$ is the number of the minimal spinors while $M$ and $N$ are the correspondences in spacetime.

The following table covers the detail of the minimal spinor and the number of the minimal spinors in diverse dimensions.
 
 \begin{center}
Minimal spinor components \& Supersymmetries in diverse dimensions\\
\vskip 0.1cm
\begin{tabular}{|c|c|c|}
\hline
Dimension (D or d) & Minimal Spinor (M or m) & SUSY (N or n)\\
\hline\hline
11 & 32 & 1\\
\hline
10 & 16  & 2, 1\\
 \hline
9 & 16 & 2, 1\\
\hline
8&16& 2, 1\\

\hline
7& 16 &2, 1\\

\hline
6& 8&4, 3, 2, 1\\

\hline
5&8&4, 3, 2, 1\\

\hline
4&4&8,  $\cdots$,  1\\

\hline
3&2&16,$ \cdots$, 1\\
 \hline
2& 1& 32, $\cdots$, 1\\
\hline
\end{tabular}
\end{center}
Worldvolume SUSY gives
\be
N_B = N_F \Rightarrow D - d =  \frac{1}{2} m n = \frac{1}{4} M N.
\ee
There are 8 solutions all with $N = 1$ when $d > 2$ (Note that $D_{\rm max} = 11$ due to $M \ge 64$ when $D \ge 12$ and $d_{\rm max} = 6$ due to $m \ge 16$ when $d \ge 7$).

The special $d = 2$ case: left and right modes, independent of each other, can be treated separately.  If both $N_B$ and $N_F$ are the sum of the left and right modes, $N_B = N_F$ gives additional four solutions all with $N = 2$ in $D = 3, 4, 6, 10$ (or 8 solutions if IIA and IIB are treated separately).

If only one mover matching is required,
\be
D - 2 = mn = \frac{1}{2} M N,
\ee
we have another four solutions all with $N = 1$ in $D = 3, 4, 6, 10$.

All these solutions give the old-brane-scan as given earlier and we give it here again for convenience (Figure 6).

 \begin{figure}[!hbp]
\begin{center}
\includegraphics[scale= 1.0]{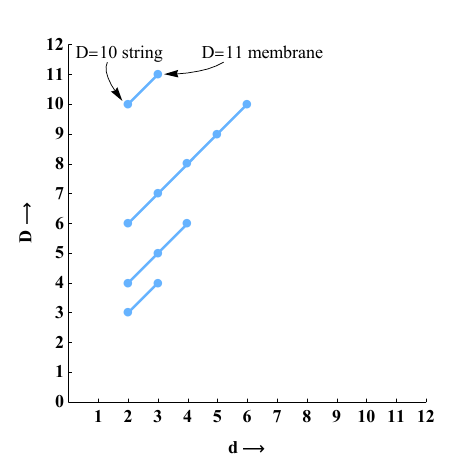}
\end{center}
\caption{The old brane scan}\label{oldbs1}
\end{figure}
 Now if the worldvolume vectors (or tensors) are also considered, the bosonic DOF are
 \be
 N_B = D - d + (d - 2) = D  - 2  = N_F = \frac{1}{2} m n = \frac{1}{4} M N,
 \ee
 where in addition to $D - d$ scalars, we have also a worldvolume massless vector in d dimensions which gives $d - 2$ physical polarizations, therefore giving a total bosonic $N_{B} = D - d  + d - 2 = D - 2$.
 A few additional facts should be noticed before proceeding.  A vector supermultiplet exists only in $4 \le d \le 10$. In $d = 3$, a vector is dual to a scalar while in $d = 2$ a vector has no propagating degree of freedom (DOF). A chiral (2, 0) tensor supermultiplet ($B^-_{\mu\nu}, \lambda^I, \phi^{[IJ]}$) exists in $d = 6$ for IIA NSNS 5-brane in D = 10 and M5 brane in D = 11. The (anti) self-dual tensor counts only 3 physical (on-shell) DOF  while the index $I = 1, 2, 3, 4$ runs in the fundamental representation of USp(4). Here $\phi^{[IJ]}$ with $I, J$ antisymmetric counts 5 scalars in this multiplet since we have the following traceless condition
 \be
 \Omega_{IJ} \phi^{[IJ]} = 0,
 \ee
 with $\Omega$ the invariant tensor of $USp(4)$ which can be chosen as
 \be
 \Omega = \left(\begin{array}{cc}
 0 & \mathbb{I}_{2 \times 2}\\
 - \mathbb{I}_{2 \times 2} & 0
 \end{array}\right).
 \ee
In the case of M5, the 5 scalars are the 5 translational modes $y^m$ with $m = 6, 7, \cdots 10$.   Due to that $SO(5) \cong USp(4)$, they can also be grouped under USp(4) as $\phi^{[IJ]}$ in the above tensor multiplet.  The 4 spinors are the symplectic Majorana-Weyl ones since in d = 6 with the symplectic USp(4), we can have the symplectic Majorana as follows
\be
\lambda_{I}^{*} = (\lambda^{I})^{*} = \Omega_{IJ} B \lambda^{J},
\ee
where the B matrix is related to the charge conjugate matrix $C = C^{T}$ as $C = B \gamma^{0}$ with $\gamma^{0}$ the 6d Dirac matrix and has the following properties
\be
B^{+} B = \mathbb{I}, \quad B^{T} = - B.
\ee
So for the M5, we have
\be
N_{B} = D - d + 3  = 11 - 6 + 3 = 8 = N_{F} = \frac{1}{2} m n = \frac{1}{4} M N,
\ee
still holds since $m = 4, n = 4$ or $M = 32, N = 1$.

\begin{figure}[t]
\begin{center}
\includegraphics[scale= 0.4]{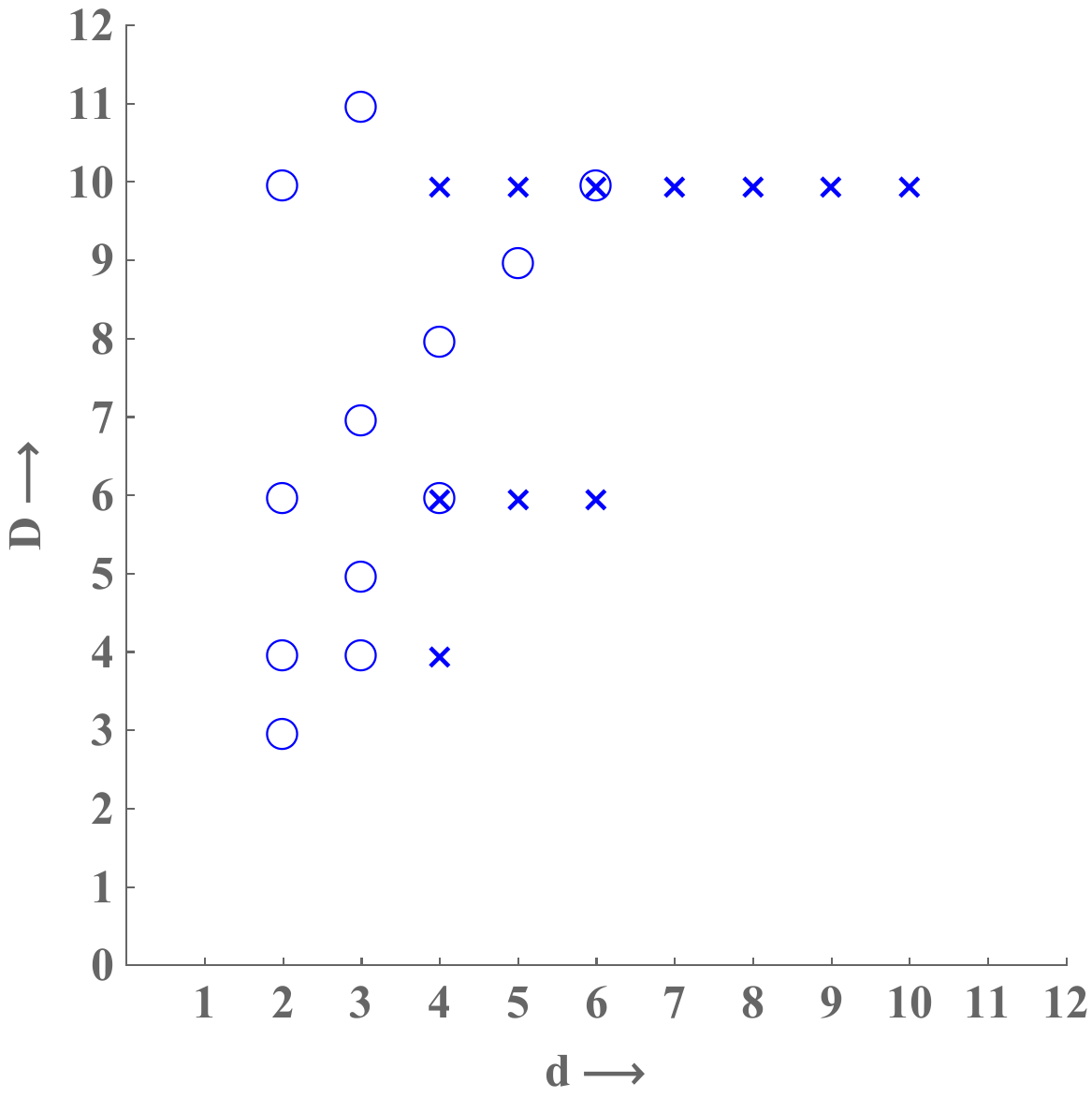}
\end{center}
\caption{The new brane scan.  All possible $d \ge 2$ scalar supermultiplets, denoted by circles, and vector supermultiplets, denoted by crosses, according to $D = d + $ the number of scalars.  }\label{newbs1}
\end{figure}

Given the above, we have the new brane-scan \cite{Duff:1992hu} as given before and is given here again as Figure \ref{newbs1}.

\medskip
\noindent
{\sl 10D 1/2 BPS p-brane zero modes (IIA \& IIB)}
\medskip

 Here we give a concrete counting of 10D 1/2 BPS p-brane zero modes. In general, the broken translation symmetries give rise to the translation zero modes while the broken SUSY gives the fermionic zero modes.

\medskip
\noindent
1/2 BPS F-string:
\medskip

8  translation modes $X^m$ with $m = 2, 3, \cdots, 9$; matching with 8  fermionic modes (16 broken SUSY generators and half of those contribute to the on-shell zero modes).  We have here a scalar supermultiplet ($\phi^I, \lambda^I$) with $I = 1, 2, \cdots, 8$ and SO(8)  R-symmetry.

\medskip
\noindent
1/2 BPS NSNS 5-brane (IIA): 
\medskip

We have 4  translation zero modes $X^m$ with $m = 6, 7, 8, 9$ and one extra scalar from the field fluctuation for the NSNS 5-brane as well as 3 tensor zero modes $B^-_{\mu\nu}$ with its field strength anti self-dual also from the fluctuations, giving a total of $8_B$ bosonic DOF (see for example \cite{Callan:1991ky}), which match with $8_F$. We have here a chiral (2, 0) tensor supermultiplet ($B^-_{\mu\nu}, \lambda^I, \phi^{[IJ]}$) with the R-symmetry USp(4) discussed earlier.

\medskip
\noindent
1/2 BPS NSNS 5-brane (IIB):
\medskip

We have 4 translation zero modes $Y^m$ ($m = 1, 2, 3, 4$), 4 zero modes coming from a (1 + 5) vector, giving a non-chiral (1, 1) 6d vector multiplet ($A_\mu, \phi^I, \lambda^I$) with $I = 1, 2, 3 , 4$ the vector index under SO$_R$ (4). Here the 4 spinors $\lambda^I$ are pseudo-Majorana -Weyl one, each counting 4 components and 2 on-shell DOF (see also \cite{Callan:1991ky}).

\medskip
\noindent
1/2 BPS Dp brane:

\medskip

We have $10 - d$ translation zero modes $X^m$ with $m = d, d + 1, \cdots, 9$ and a d-dimensional vector counting $d - 2$ zero modes (This vector can be found in a similar fashion following \cite{Duff:1991pea, Callan:1991ky} from fluctuations around the p-brane background configuration), giving a total of $10 - 2 = 8$ bosonic zero modes, matching with the 8 fermionic zero modes (half of the broken SUSY generators). We have a vector supermultiplet on the brane ($A_\mu, \phi^I, \lambda^I$) with $I = 1, 2, \cdots, 10 - d$ and the R-symmetry SO (10 - d). Here $\lambda^I$ are  Majorana or pseudo or symplectic Majorana and can further be Weyl if the worldvolume dimension is 2 or 6 or 10.

\section{Non-SUSY \& Non-BPS p-branes in diverse dimensions}

We consider in this section the Non-SUSY and Non-BPS p-brane configurations in the simplest setting in the sense consisting of only one type\footnote{In other words, we don't consider a Non-SUSY configuration which includes different types of branes, for example, different p's or those with D branes and NSNS branes.} (including anti ) p-branes put on top of each other.  We have in general two kinds of such configurations which do not preserve any SUSY.

One is the black version of the BPS p-branes discussed in the above for which I don't have intension to discuss here\footnote{These can be found in any of the references \cite{Horowitz:1991cd} in 10D and in diverse dimensions \cite{Duff:1993ye, Duff:1994an, Duff:1996hp}} except for the following brief mentioning. The general feature for the metric is, for BPS p-brane,
\be
d s^{2} = e^{2A (r)} \left(- d t^{2} + d x_{1}^{2} + \cdots + d x^{2}_{p} \right) + e^{2 B (r)} \left( d r^{2} + r^{2} d\Omega^{2}_{\tilde d + 1} \right),
\ee
while for the corresponding black p-brane,
\be
d s^{2} =  e^{2A (r)} \left(-  f (r) d t^{2} + d x_{1}^{2} + \cdots + d x^{2}_{p} \right) + e^{2 B (r)} \left(  f^{- 1} (r) d r^{2} + r^{2} d\Omega_{\tilde d + 1}^{2} \right),
\ee
and there is an event-horizon occurring at
\be
f (r) = 1 - \frac{K_{p}}{r^{\tilde d}} = 0 \to r_{+} = K^{1/\tilde d}_{p}.
\ee
In the above $\tilde d = D - d - 2$.

However, there is the other kind of the non-SUSY p-branes whose ansatz for the metric looks no different from the BPS one, i.e.,
\be\label{non-susy-m}
d s^{2} = e^{2A (r)} \left(- d t^{2} + d x_{1}^{2} + \cdots + d x^{2}_{p} \right) + e^{2 B (r)} \left( d r^{2} + r^{2} d\Omega_{\tilde d + 1}^{2} \right),
\ee
and the ansatz for the other fields remain also the same. The resulting p-brane is not a BPS one or does not respect any SUSY simply because
\be
d \,A (r) + \tilde d\, B (r) = \ln G (r) \neq 0,
\ee
if one recalls the condition given in (\ref{SUSY1}) and the discussion thereafter for p-brane preserving SUSY.  This kind of non-susy p-brane solutions were first considered by Zhou and Zhu in \cite{Zhou:1999nm} for pure solution purpose but the physical meaning of these configurations as representing a brane/anti brane system or non-BPS branes was later noticed by  Brax et al \cite{Brax:2000cf} and further studied  by the present author and his collaborator Roy \cite{Lu:2004dp, Lu:2004xi}.

 A more general such non-SUSY p-brane solutions were given later by the present author and his collaborator Roy in \cite{Lu:2004ms}.

 \begin{figure}[t]
\begin{center}
\includegraphics[scale= 1.0]{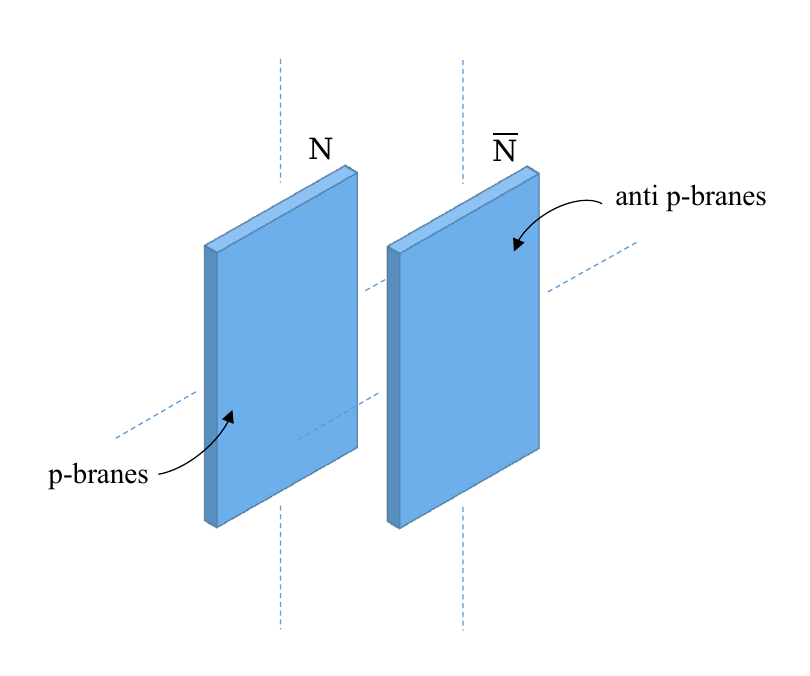}
\end{center}
\caption{Brane-Anti Brane configuration}\label{non-susy-p-brane}
\end{figure}

There is mounting evidence in support of this interpretation for this kind non-susy p-brane solutions (see Figure \ref{non-susy-p-brane}). There is an attractive force acting between $N$  p-branes and $\bar N$ anti p-branes when they are placed parallel to each other. When their separation is on the order of string length $l_s$, annihilation between the branes and the anti branes begins to occur in reality.  But on the supergravity approximation, such an annihilation process will not manifest to the observer outside the event horizon and instead when the $N$ p-branes and the $\bar N$ anti p-branes are put on top of each other, the apparent spherical symmetry along the transverse directions will give rise to a static configuration due to the Birkhoff's theorem.

The other way to see this is to show that on the supergravity approximation, the attractive force between the $N$ p-brane \& $\bar N$ p-branes vanishes when they are put on top of each other  \cite{Lu:2006rb}.   As expected and given above, these non-susy p-branes have also the symmetry
  \be
  P_{d} \times SO (D - p - 1).
  \ee

 Unlike the SUSY case, the brane source in the present context will not be useful  or helpful in finding solution and instead can make things even more complicated.  As discussed earlier, if using the dual formalism and  making ansatz on the dual field strength, we can forget about the brane source altogether. We will adopt this in what follows by making an ansatz on the dual $(1 + \tilde d)$-form field strength
 \be\label{formf-ansatz}
 F_{1 + \tilde d} = b {\rm Vol} (\Omega_{D - p - 2}),
 \ee
 where $b$ is just a constant flux and ${\rm Vol} (\Omega_{D - p - 2})$ is the volume-form on the unit $S^{D - p - 2}$, therefore this brane system carries a net ``magnetic-like" charge.  In other words, we have
 \be
 F_{1 + \tilde d} = b \sqrt{\det \bar g_{\bar a \bar b}} \, d\theta^{1}\wedge \cdots \wedge d \theta^{1 + \tilde d},
 \ee
 where $\bar g_{\bar a \bar b}$ is the metric on the unit $S^{1 + \tilde d}$ sphere and $\theta^{\bar a}$ are the angle coordinates defined on the sphere. So we have the following only non-vanishing sum
 \bea
 F_{\bar a M_{1}\cdots M_{d}} F^{\bar b M_{1} \cdots M_{d}} &=& d! b^{2} \delta_{\bar a}\,^{\bar b} r^{- 2 (1 + \tilde d)} e^{- 2 (1 + \tilde d) B} (\det \bar g_{\bar a\bar b})^{-1} \det \bar g_{\bar a\bar b} \nn
 &=& d! \frac{b^{2}}{r^{2 (1 + \tilde d)}} e^{- 2 (1 + \tilde d)B } \delta_{\bar a}\,^{\bar b},
 \eea
 which gives
 \be
 F^{2}_{1 + \tilde d } = (1 + \tilde d)! \frac{b^{2}}{r^{2 (1 + \tilde d)}} e^{- 2 (1 + \tilde d) B}.
 \ee 
 Since we make the ansatz for the dual field strength as given above,  the field strength appearing in the bulk spacetime action (\ref{bulkA}) should be this one and therefore the dilation coupling is the dual one $\alpha (\tilde d) = - \alpha (d)$.  Also because of this, we don't need to have the source brane present and this makes all the EOM absence of the source. Apart from these, all the other things in EOM should remain the same as the BPS case considered in the previous sections.  For convenience, we still write the EOM for all the bulk fields in the following:
  \bea\label{non-susy-EOM0}
  R_{MN} - \frac{1}{2} \partial_{M}\bar\phi \partial_{N} \bar\phi \qquad\qquad \qquad\qquad\qquad\qquad\qquad\qquad\qquad\quad\nn
  - \frac{e^{\alpha (d) \bar\phi}}{2 \tilde d !} \left[F_{MM_{1} \cdots \tilde d} F_{N}\,^{M_{1} \cdots M_{\tilde d}}
   - \frac{\tilde d}{ (\tilde d + 1)(D - 2)} g_{MN} F^{2}_{\tilde d + 1} \right] & = & 0,\label{non-susy-EOM01}\\
  \partial_{M}\left(\sqrt{- g}\, e^{\alpha (d) \bar\phi} F^{MM_{1} \cdots M_{\tilde d}} \right) &=& 0, \label{non-susy-EOM02}\\
  \frac{1}{\sqrt{- g}} \partial_{M} \left(\sqrt{- g} g^{MN} \partial_{N} \bar\phi \right) - \frac{\alpha (d)}{2 (\tilde d + 1)!}\, e^{\alpha (d) \bar\phi} F^{2}_{\tilde d + 1} &=& 0.\label{non-susy-EOM03}
  \eea
  Since we are interesting the Non-SUSY solutions, as discussed above, we here have the ansatz
  \be\label{G}
  d A + \tilde d B = \ln G (r).
  \ee
  Given the ansatz (\ref{non-susy-m}) for the metric, we have (i.e., those given in (\ref{ricci-tensor}) in spherical coordinates) 
  \bea\label{curvature}
  R_{rr} &=& - d \left( A'' + A'^{2} - A' B' \right) - (\tilde d + 1) \left(B'' + \frac{B'}{r} \right),\nn
  R_{xx}& =& - R_{tt} =  - e^{2 (A - B) } \left(A'' + \tilde d A' B' + d A'^{2} + (\tilde d + 1) \frac{A'}{r}\right),\nn
  R_{\bar a \bar b} &= & -  r^{2} \left[ B'' + \tilde d B'^{2} + \frac{2 \tilde d + 1}{r} B' + d A' \left(B' + \frac{1}{r} \right)\right] \bar g_{\bar a \bar b},
  \eea
where $\bar a, \bar b$ are the indices for the transverse spherical (angular) coordinates and $\bar g_{\bar a \bar b}$ is the metric for the unit $(\tilde d + 1)$-sphere. Here $x$ stands for $x^{i}$ with $i = 1, \cdots, p$ ($d =  1 + p$). The `prime' here denotes the derivative with respect to $r$. Note that given the ansatz for the $(1 + \tilde d)$-form field strength (\ref{formf-ansatz}), its corresponding EOM (\ref{non-susy-EOM02}) satisfies automatically. We have the EOM as 
   \bea\label{non-susy-EOM1}
   B'' + \frac{G''}{G} - \frac{G'^{2}}{G^{2}} + \frac{1}{d}\left(\frac{G'}{G} - \tilde d B'\right)^{2}  + \tilde d\, B'^{2} - \frac{G'}{G} B' &\,&\nn
   + \frac{\tilde d + 1}{r} B' + \frac{1}{2} \bar\phi'^{2} - \frac{\tilde d\, b^{2}}{2 (D - 2)} \frac{e^{2 d A + \alpha \bar\phi}}{G^{2} r^{2 (\tilde d + 1)}} & = & 0,\label{non-susy-EOM11}\\
   A'' + \frac{\tilde d + 1}{r} A' + \frac{G'}{G} A' - \frac{ \tilde d\, b^{2}}{2 (D - 2)}    \frac{e^{2 d A + \alpha \bar\phi}}{G^{2} r^{2 (\tilde d + 1)}}    & = & 0, \label{non-susy-EOM12}\\
   B'' + \frac{\tilde d + 1}{r} B' + \frac{G'}{G}\left(B' + \frac{1}{r}\right) + \frac{ d\, b^{2}}{2 (D - 2)}  \frac{e^{2 d A + \alpha \bar\phi}}{G^{2} r^{2 (\tilde d + 1)}} & = & 0,\label{non-susy-EOM13}\\
   \bar\phi'' + \left(\frac{\tilde d + 1}{r} + \frac{G'}{G}\right) \bar\phi' - \frac{\alpha b^{2}}{2}  \frac{e^{2 d A + \alpha \bar\phi}}{G^{2} r^{2 (\tilde d + 1)}} & =& 0.\label{non-susy-EOM14}
   \eea
  From (\ref{non-susy-EOM12}) and (\ref{non-susy-EOM13}) and using $d A + \tilde d B = \ln G$, we have
   \be\label{G-EOM}
  G'' + \frac{2 \tilde d + 1}{r} G' = 0.
  \ee
Noting $G(r \to \infty) \to 1$ from (\ref{G}) due to $A (r \to \infty) \to 0$ and $B (r \to \infty) \to 0$, we have the following two solutions
  \be\label{G-solution}
  G_{-} =  1 - \left(\frac{\omega}{r}\right)^{2 \tilde d}, \qquad G_{+} = 1 + \left(\frac{\tilde \omega}{r}\right)^{2 \tilde d},
  \ee
  where both $\omega$ and $\tilde \omega$ are taken as positively real.
 
  In what follows, we focus on 
  \be\label{G-}
  G_{-} (r) = \left(1 + \frac{\omega^{\tilde d}}{r^{\tilde d}}\right) \left(1 - \frac{\omega^{\tilde d}}{r^{\tilde d}}\right) \equiv H (r) \tilde H (r),
  \ee
  while a detail discussion  for the case of $G_{+}$ is referred to \cite{Lu:2004ms}.   In the above
  \be
  H (r) = 1 + \frac{\omega^{\tilde d}}{r^{\tilde d}}, \qquad \tilde H (r) = 1 - \frac{\omega^{\tilde d}}{r^{\tilde d}}.
  \ee
  With this $G_{-} (r)$, we have, from (\ref{non-susy-EOM12}) and (\ref{non-susy-EOM14}), 
  \be
  \left(\bar\phi - \frac{\alpha (D - 2)}{\tilde d} A\right)'' + \left(\frac{\tilde d + 1}{r} + \frac{G'_{-}}{G_{-}}\right) \left(\bar\phi - \frac{\alpha (D - 2)}{\tilde d} A\right)' = 0,
  \ee
   which has the solution
  \be
  \bar\phi = \frac{\alpha (D - 2) }{\tilde d} A + \delta \ln \frac{H}{\tilde H},
  \ee
  with $\delta$ a parameter.  We then have
  \be
  e^{2 d A + \alpha \bar\phi} = \left(\frac{H}{\tilde H}\right)^{\alpha \delta} e^{\chi A},
  \ee
  where
  \be
  \chi = 2 d + \frac{\alpha^{2} (D - 2)}{\tilde d}.
  \ee
  We have then from (\ref{non-susy-EOM12}) for $A$  as
  \be\label{A}
  A'' + \left(\frac{\tilde d + 1}{r} + \frac{G'_{-}}{G_{-}} \right) A' - \frac{\tilde d\, b^{2}}{2 (D - 2)} \frac{e^{\chi A}}{r^{2 (\tilde d + 1)}} \frac{H^{\alpha \delta - 2}}{\tilde H^{\alpha \delta + 2}} = 0.
  \ee
 To have a solution of the above, we make the following ansatz
  \be\label{A-ansatz}
  e^{A} = \left[\left(\frac{H}{\tilde H}\right)^{\bar\alpha} \cosh^{2} \theta - \left(\frac{\tilde H}{H}\right)^{\bar\beta} \sinh^{2} \theta \right]^{\gamma} \equiv F^{\gamma},
  \ee
  where
  \be
  F =  \left(\frac{H}{\tilde H}\right)^{\bar\alpha} \cosh^{2} \theta - \left(\frac{\tilde H}{H}\right)^{\bar\beta} \sinh^{2} \theta,
  \ee
  and the parameters $\bar\alpha, \bar\beta, \theta$ and $\gamma$ are to be determined shortly.  With this $A$, we can obtain $B$ from $d A + \tilde d B = \ln G_{-}$.  
  Plug this $A$ along with $G_{-}$ given in (\ref{G-}) into (\ref{A}), we have
  \be\label{A-1}
 \left( \gamma \tilde d\, \omega^{2 \tilde d} (\bar\alpha + \bar \beta)^{2} \sinh^{2} 2 \theta \right)\, H^{\bar\alpha - \bar\beta} \tilde H^{\bar\beta - \bar\alpha} F^{-2} + \frac{b^{2}}{2 (D - 2)} H^{\alpha \delta} \tilde H^{- \alpha \delta} F^{\chi \gamma} = 0.
 \ee
 In having the above, we need to pay attention to the following. From (\ref{A-ansatz}), we have
 \be
 A = \gamma \ln F, \quad A' = \gamma \frac{F'}{F}, \quad A'' = \gamma \left(\frac{F''}{F} - \frac{F'^{2}}{F^{2}}\right),
 \ee 
 then in terms of $F,\, F'$ and $F''$,  (\ref{A}) becomes 
 \be\label{A-2}
 \gamma \left[  \left(\frac{F''}{F} - \frac{F'^{2}}{F^{2}}\right) +  \left(\frac{\tilde d + 1}{r} + \frac{G'_{-}}{G_{-}} \right) \frac{F'}{F}\right] - \frac{\tilde d\, b^{2}}{2 (D - 2)} \frac{F^{\chi \gamma}}{r^{2 (\tilde d + 1)}} \frac{H^{\alpha \delta - 2}}{\tilde H^{\alpha \delta + 2}} = 0. 
 \ee
 In order to obtain the first term on the left side of (\ref{A-1}) with ease, the order regrouping terms in the above will be important.  We first combine the following terms with some efforts to give
 \bea
 F'' +  \left(\frac{\tilde d + 1}{r} + \frac{G'_{-}}{G_{-}} \right) F' &=& \bar\alpha^{2} \left(\frac{H}{\tilde H}\right)^{\bar\alpha} \left(\frac{H'}{H} - \frac{\tilde H'}{\tilde H}\right)^{2} \cosh^{2}\theta \nn
 &\,& - \bar\beta^{2} \left(\frac{\tilde H}{H}\right)^{\bar\beta}  \left(\frac{H'}{H} - \frac{\tilde H'}{\tilde H}\right)^{2} \sinh^{2}\theta,
 \eea 
 then we can have
 \bea
 F\left[ F'' +  \left(\frac{\tilde d + 1}{r} + \frac{G'_{-}}{G_{-}} \right) F' \right] - F'^{2} &=& - \left(\frac{H'}{H} - \frac{\tilde H'}{\tilde H}\right)^{2} \frac{\left(\bar\alpha + \bar\beta\right)^{2}}{4} \left(\frac{H}{\tilde H}\right)^{\bar\alpha - \bar\beta} \sinh^{2} 2 \theta \quad \nn
 &=& - \frac{\left(\bar\alpha + \bar\beta\right)^{2}\, \tilde d^{2} \, \omega^{2\tilde d}}{r^{2(1 + \tilde d)}} H^{\bar\alpha - \bar\beta - 2}  \tilde H^{\bar\beta - \bar\alpha - 2} \sinh^{2} 2\theta.
 \eea
  With this, (\ref{A-2}) becomes (\ref{A-1}).   To have (\ref{A-1}) hold in general, we need the parameters to satisfy the following relations
 \bea\label{parameter-relation}
 \gamma &=& - \frac{2}{\chi}, \qquad \bar\alpha - \bar \beta = \alpha \delta, \nn
 b & =  & \sqrt{\frac{4 \tilde d \, (D - 2)}{\chi}} \, (\bar\alpha + \bar \beta) \omega^{\tilde d} \sinh \theta.
 \eea
 We have left one equation (\ref{non-susy-EOM11}) unsolved. Plug $B, A, \bar\phi$ along with $G_{-}$ into this equation, we find a further constraint on the parameters as given below
 \be
 \frac{1}{2} \delta^{2} + \frac{2 \bar\alpha (\bar\alpha - \alpha \delta)(D - 2)}{\tilde d \, \chi} = \frac{\tilde d + 1}{\tilde d}.
 \ee
 This can be solved, when combined with the relations given in (\ref{parameter-relation}), to give      
  \bea\label{parameter-solution}
  \gamma &=& - \frac{2}{\chi},\nn
  \bar\alpha &=& \sqrt{\frac{\chi (\tilde d + 1)}{2 (D - 2)} - \frac{\delta^{2}}{4} \left(\frac{ \tilde d \chi}{D - 2} -\alpha^{2}\right)} + \frac{\alpha \delta}{2},\nn
  \bar\beta &=&  \sqrt{\frac{\chi (\tilde d + 1)}{2 (D - 2)} - \frac{\delta^{2}}{4} \left(\frac{ \tilde d \chi }{D - 2} -\alpha^{2}\right)} - \frac{\alpha \delta}{2},\nn
  b &=& \sqrt{\frac{4 \tilde d (D - 2)}{\chi}} (\bar \alpha + \bar\beta) \omega^{\tilde d} \sinh\theta.
  \eea
  So this solution has three parameters $\delta, \omega$ and $\theta$.  Note that
  \be
  e^{2A} = F^{- \frac{4}{\chi}} ,  \qquad e^{2 B} = \left(H \tilde H\right)^{\frac{2}{\tilde d}} \, F^{\frac{4 d}{\tilde d \chi}}.
  \ee
  The complete non-susy p-brane solution is
  \bea\label{non-susy-p-brane-solution}
  d s^{2} &= & F^{- \frac{4}{\chi}} \left(- d t^{2} + d x^{2}_{1} + \cdots d x^{2}_{p} \right) + \left(H\tilde H\right)^{\frac{2}{\tilde d}} F^{\frac{4 d}{ \tilde d \chi}} \left(d r^{2} + r^{2} d\Omega^{2}_{\tilde d + 1} \right),\nn
  e^{2\bar\phi} &=& F^{- \frac{4 \alpha (D - 2)}{ \tilde d \chi}} \left(\frac{H}{\tilde H}\right)^{2 \delta}, \nn
  F_{\tilde d + 1} &=& b {\rm Vol} (\Omega_{\tilde d + 1}).
  \eea
  This is a three-parameter ($\omega, \theta, \delta$) solution.
  
  \medskip
  \noindent
  {\bf Chargeless Solution:  $b = 0$}
  \medskip

1) $\bar\alpha + \bar \beta = 0$ or $\delta^{2} = \frac{2 \chi (\tilde d + 1)}{(D - 2) ( \tilde d\chi - \alpha^{2} (D - 2))}$. We have a non-trivial configuration with $b = 0$, implying $N = \bar N$, the number of branes and that of the anti brane are the same.

\medskip

2) $\omega = 0$. This solution is trivial Minkowski spacetime, preserving all SUSY ($N = \bar N = 0$).

\medskip

3) $\theta = 0$. Now $F = \left(\frac{H}{\tilde H}\right)^{\bar\alpha}$ and this solution is still a non-trivial one, giving $N = \bar N \neq 0$.

\medskip

\noindent
{\bf BPS p-brane limit:}
\medskip
  \bea
  (\bar\alpha + \bar\beta) \omega^{\tilde d} &\to& \epsilon \bar\omega^{\tilde d}, \nn
    \sinh 2 \theta &\to& \epsilon^{- 1},
    \eea
    with $\epsilon \to 0$ ($\omega \to 0$) while keeping $\bar\omega$ fixed.  We then have
    \bea
    H (r) &\to& 1, \qquad \tilde H (r)  \to 1,\nn
    F(r) &\to & 1 + \frac{\left[(\bar\alpha + \bar\beta) \cosh2\theta + (\bar\alpha - \bar\beta)\right]\omega^{\tilde d}}{r^{\tilde d}} \to 1 + \frac{\bar\omega^{\tilde d}}{r^{\tilde d}} = \bar H (r).
    \eea
    The non-susy configurations become the corresponding BPS p-brane
    \bea
    d s^{2} &=& \bar H^{- \frac{4}{\chi}} \left(- d t^{2} + d x^{2}_{1} + \cdots d x^{2}_{p} \right) + \bar H^{\frac{4 d}{ \tilde d \chi}} \left(d r^{2} + r^{2} d\Omega^{2}_{\tilde d + 1} \right),\nn
  e^{2\bar\phi} &=& \bar H^{- \frac{4 \alpha (D - 2)}{\tilde d\chi}}, \nn
  F_{\tilde d + 1} &=& b {\rm Vol} (\Omega_{\tilde d + 1}).
  \eea
   Here $e^{- C} = \bar H$ and this gives
   \be
   A = \frac{2}{\chi} C, \quad B = - \frac{ 2 d}{\tilde d \chi}C,
   \ee
with $ d A + \tilde d B = 0$, therefore SUSY preserved.   This implies either
     \be
     N \to 0, \quad {\rm or}\quad \bar N  \to 0,
     \ee
     but keeping $b$ fixed. The previous chargeless case with $\omega = 0 \to N = \bar N  = 0$ $ \to \bar\omega = 0$. 
 
For $b \neq 0$ non-susy p-brane solution, it is interpreted as representing $N$ (p-brane) - $\bar N$ (anti p-brane) system (Note p = even in II A  and p = odd in IIB).
  
Howver, for $b = 0$ non-trivial solutions: p = even, brane-anti brane pair in IIA but non-BPS in IIB while p = odd, brane-anti brane in IIB but non-BPS in IIA \cite{Sen:1998sm, Sen:1999mg}.

Evidence for these interpretations has been given in a series of works on tachyon condensation picture based on open string which can be realized from the above solutions (see, \cite{Lu:2004dp, Bai:2005jr,Lu:2004xi}). The descent relations relating brane-anti brane pair, non-BPS brane \& BPS brane from open string can also be realized with some delocalized non-susy p-brane solution in a similar fashion \cite{Lu:2005ju}.

Some non-perturbative issues regarding brane-anti brane pair or non-BPS brane can also be addressed using the supergravity approach \cite{Lu:2005jc} but here caution needs to be paid about the validity of the solutions. Various puzzling issues regarding supergravity approach have also been addressed in the work \cite{Lu:2006rb}.

\section*{Acknowledgments}
The author would like to thank Nan Zhang for help in drawing the figures in this paper and acknowledges the support by grants from the NNSF of China with Grant No: 12275264 and 12247103.


\begin{thebibliography}{99}
\bibitem{lu1}
For examples, for Type I supergravity plus super Yang-Mills with gauge group SO(32), this theory by itself has nothing to do with the string coupling. If one chooses to identify it with the low energy effective theory describing  the massless modes of either the perturbative Type I superstring or the perturbative heterotic string, the gravity coupling can now be related to the corresponding string coupling. Given that the two strings are related by strong-weak coupling, it  must  therefore imply that the supergravity plus the super Yang-Mills itself as the underlying low energy effective theory is good for both weak and strong couplings, which is consistent with that this low energy theory has nothing to do with the string coupling and is good for any string coupling in the low energy limit. The same discussion applies to Type IIB supergravity.  It will be a bit subtle about the relation between the 10D Type IIA supergravity and the 11D supergravity. The spacetime dimensionality difference between the two makes the labelling of the fields different in the two theories. This difference can be attributed to the difference between their respective low energy scales since we are talking here a low energy effective theory whose actual form depends on the relevant energy scale as in every effective field description. For convenience, let us introduce a dimensional regulator by compactifying one of the spatial dimensions in 11D to a big circle with a physical radius $R$. Then the 10D Newtonian constant $G_{10}$ is related to 11D one $G_{11}$ via $G_{11} = 2\pi R\, G_{10}$. So the 11D Planck scale $M_{11, P} \sim G^{- 1/9}_{11} \sim R^{- 1/9} G^{- 1/9}_{10} \sim  R^{- 1/9} M^{8/9}_{10, P}$ with $M_{10, P}$ the 10D Planck scale. So we have $M_{11, P} \sim (M_{10, P} R)^{- 1/9} M_{10, P}$.  It is clear that we cannot set $M_{11,P} \sim M_{10, P}$ since if so, we would have $(M_{10, P} R)^{- 1/9} \sim {\cal O} (1)$ for arbitrary large $R$ when $M_{10, P}$ is fixed and this is obviously impossible. In other words, we have two different energy scales and the actual low energy effective theory is related to the lower one and this is in turn related to the spacetime dimensionality even though for both cases the bosonic and fermionic on-shell degrees of freedom are all given by $128_{B} + 128_{F}$. For example, if $M_{11, P} > M_{10, P}$, the relevant low energy scale is $M_{10, P}$  and the low energy effective theory is 10D supergravity for energies smaller than $ M_{10, P}$.  This is also consistent with our usual understanding in that since  now $(M_{10, P} R)^{-1/9} > 1$ from the above relation between $M_{11, P}$ and $M_{10, P}$, we have then $M_{10, P} < 1/R$ with $1/R$ the scale for the KK modes,  i.e., the 10D Planck scale being the smallest one among the relevant scales.  So the actual low energy scale determines how to represent the on-shell $128_{B} + 128_{F}$ in a covariant way in the corresponding dimensionality. String coupling doesn't enter this game explicitly but the low energy scale picture just discussed is consistent with our stringy one in the following sense. In terms of string parameters, we know that $M_{11, P} = g_{s}^{-1/3} M_{s}$ while $M_{10, P} =  g_{s}^{- 1/4} M_{s}$ with $g_{s}$ the string coupling and $M_{s}$ the string scale.  Note also $R = g_{s} M^{-1}_{s}$.  When $g_{s} \to 0$, we are in 10D since $M_{10, P}  /M_{11, P} = g_{s}^{- 1/4} /g_{s}^{- 1/3} = g_{s}^{1/12} \to 0$. In other words, the 10D Planck scale is much less than the 11D one, so the 10D supergravity is the low energy effective one, a consistent result. While on the other hand, when $g_{s} \to \infty$, $M_{10, P}/M_{11, P} =g_{s}^{1/12} \to \infty$. This says that the 11D Planck scale is much less than 10D one. So the 11D supergravity is the low energy effective one of the underlying dynamics, again a consistent result.  By similar token, the ${\cal N} =  1$ (or Type I) supergravity plus super Yang-Mills with gauge group $E_{8} \times E_{8}$ can also be discussed, using the low energy scales, one in 10D and the other in 11D but with its structure $R^{1, 9} \times S^{1}/Z_{2}$.
\end{thebibliography}
\end{document}